\documentclass[%
 aip,
 amsmath,amssymb,
 reprint,%
]{revtex4-1}

\usepackage{graphicx}%
\usepackage{dcolumn}%
\usepackage{bm}%

\usepackage[utf8]{inputenc}
\usepackage[T1]{fontenc}
\usepackage{mathptmx}

\usepackage{xcolor}

\usepackage{array}
\newcolumntype{P}[1]{>{\centering\arraybackslash}p{#1}}

\begin{document}

\title{Pressure control using stochastic cell rescaling}

\author{Mattia Bernetti}
\affiliation{Scuola Internazionale Superiore di Studi Avanzati, Via Bonomea 265, Trieste 34136, Italy}%
\author{Giovanni Bussi}
 \email{bussi@sissa.it}
\affiliation{Scuola Internazionale Superiore di Studi Avanzati, Via Bonomea 265, Trieste 34136, Italy}%

\date{\today}%

\begin{abstract}
Molecular dynamics simulations require barostats to be performed at constant pressure. The usual recipe is to
employ the Berendsen barostat first, which displays a first-order volume relaxation efficient in equilibration
but results in incorrect volume fluctuations,
followed by a second order or Monte Carlo barostat for production runs.
In this paper, we introduce stochastic cell rescaling, a first-order barostat that samples the correct volume fluctuations 
by including a suitable noise term.
The algorithm is shown to report volume fluctuations compatible with the isobaric ensemble
and its anisotropic variant is tested on a membrane simulation.
Stochastic cell rescaling can be straightforwardly implemented in existing codes and can be used effectively both
in equilibration and in production phases.
\end{abstract}

\maketitle

\section{Introduction}

\begin{figure}
\includegraphics[width=\columnwidth]{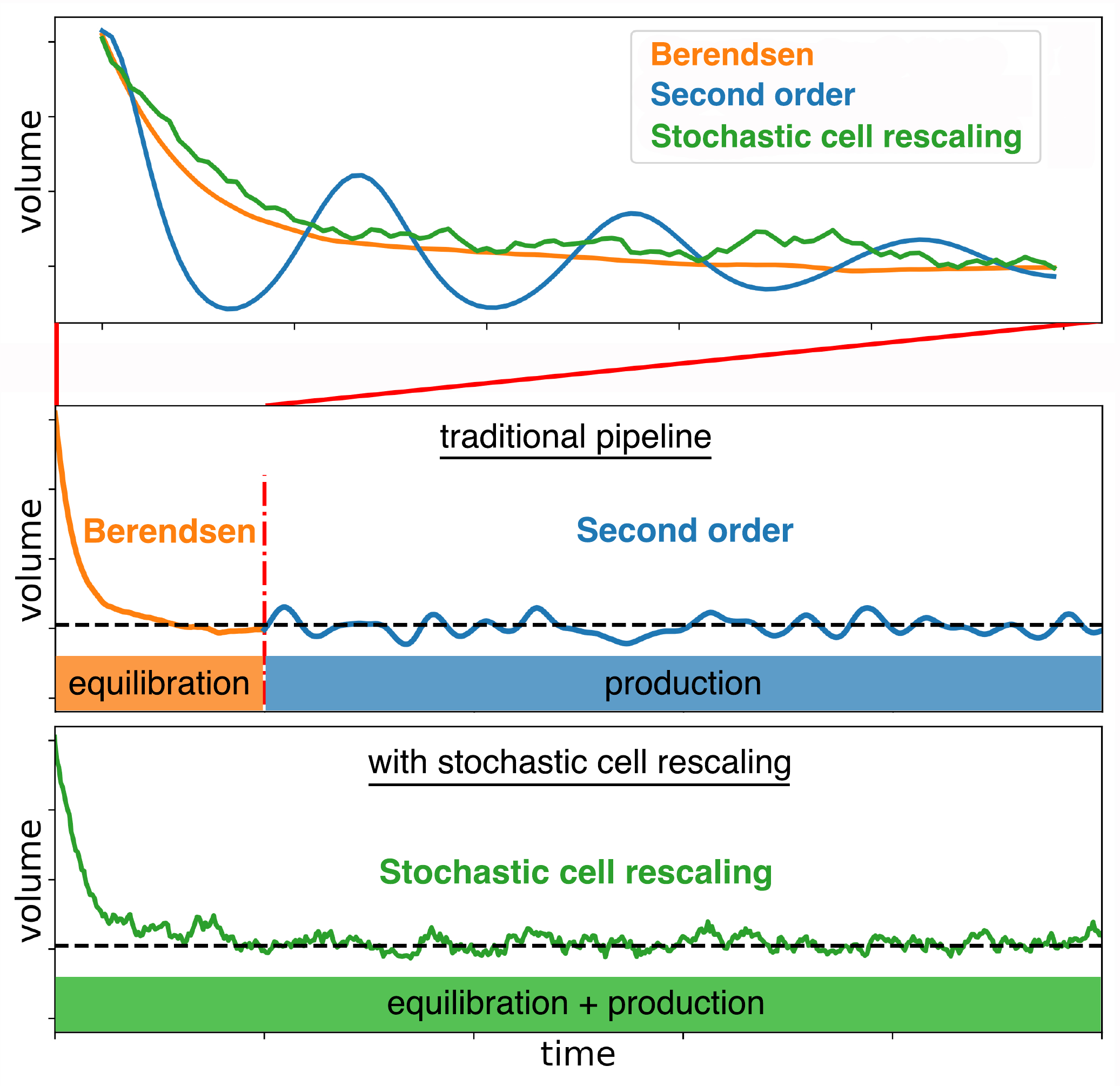}
\caption{
Graphical representation of protocols for a constant-pressure simulation.
In the traditional pipeline (middle panel), an equilibration run using a first-order Berendsen barostat\cite{berendsen1984molecular}
is followed by a production run using a second-order barostat.\cite{andersen1980molecular,parrinello1980crystal,parrinello1981polymorphic,nose1983constant,hoover1985canonical,melchionna1993hoover,martyna1994constant,zhang1995computer,martyna1996explicit,sturgeon2000symplectic,kalibaeva2003constant,marry2007trotter,bussi2009isothermal,yu2010measure,raiteri2011reactive,lippert2013accurate,feller1995constant,kolb1999optimized,quigley2004langevin,gronbech2014constant,di2015stochastic,gao2016sampling,cajahuaringa2018stochastic}
Indeed, using a second-order
barostat on a non-equilibrated system might lead to oscillations and instabilities (upper panel).
The here introduced stochastic cell rescaling algorithm relaxes straight to the correct volume
and then produces correct fluctuations (lower panel).
It can thus be used both for equilibration and production runs.
\label{fig_scheme}
}
\end{figure}

Molecular dynamics (MD) simulations can be used to characterize the dynamical properties of microscopic systems by simulating their evolution
according to the Hamilton equations of motion.\cite{frenkel2001understanding}
However, the Hamilton equations are valid only for isolated systems and need to be amended to describe the coupling with external baths.
Most common cases are thermostats and barostats. The former are used to transfer heat so as to properly control temperature. The latter are used to transfer mechanical work, thereby allowing for external pressure, stress, or surface tension to be controlled. 
Pressure control in MD was first introduced in the pioneering work of Andersen\cite{andersen1980molecular}
using an extended Lagrangian formalism.
This framework was then extended to allow periodic cells of arbitrary shapes.\cite{parrinello1980crystal,parrinello1981polymorphic}
A number of variants of these methods have been published \cite{nose1983constant,hoover1985canonical,melchionna1993hoover,martyna1994constant,zhang1995computer,martyna1996explicit,sturgeon2000symplectic,kalibaeva2003constant,marry2007trotter,bussi2009isothermal,yu2010measure,raiteri2011reactive,lippert2013accurate}
with either improved integration schemes, small modifications to control errors when the number of simulated particles is small,
or generalization to liquid interfaces.
The volume degree of freedom can also be coupled to a stochastic thermostat in a so-called Langevin piston approach.
\cite{feller1995constant,kolb1999optimized,quigley2004langevin,gronbech2014constant,di2015stochastic,gao2016sampling,cajahuaringa2018stochastic}
All the methods mentioned so far associate an inertia to the volume,
resulting in a second-order differential equation for its time evolution, which is stochastic
for Langevin piston algorithms.
An alternative approach is to use a Monte Carlo procedure to resample the volume every few steps of MD.\cite{chow1995isothermal,aaqvist2004molecular}
The Monte Carlo barostat is simpler to implement, since it does not require
the calculation of the virial, but it is sometime considered less efficient than virial-based barostats.\cite{harger2019virial}
The only barostat based on a first-order differential equation is the weak coupling or Berendsen barostat.\cite{berendsen1984molecular}
This barostat intuitively changes the volume by an increment proportional to the difference between the internal and external pressure and is very efficient
in equilibrating the system. However, it does not sample a predictable ensemble.
The usual rule of the thumb is thus to use the Berendsen barostat for equilibration, followed by either a second-order
barostat or a Monte Carlo barostat for production (see Fig.~\ref{fig_scheme}, middle panel).\cite{braun2019best}

In this work, we propose a new scheme that is based on a first-order differential equation but, at variance with the Berendsen scheme,
samples the correct isothermal-isobaric ensemble (see Fig.~\ref{fig_scheme}, lower panel).
To achieve this result, we add a suitably designed stochastic term to the Berendsen barostat,
in the spirit of what was done in the stochastic velocity rescaling algorithm\cite{bussi2007canonical} to amend the Berendsen thermostat.
The resulting algorithm can also be seen as a high-friction variant of the Langevin piston barostat.
We derive and test several possible integration schemes and provide a reference implementation for all of them
in an educational MD code.
One of the integration schemes is also implemented and tested in a modified version of the popular MD code GROMACS.\cite{abraham2015gromacs}
The algorithm produces correct volume fluctuations for a wide
range of the control parameters in a Lennard-Jones fluid and in liquid water. The method is also tested on the calculation
of solvation free energies, for which a recent paper suggested incorrect results arising from use of the Berendsen barostat,\cite{rizzi2020sampl6}
and its semi-isotropic version is tested on a membrane simulation.

\section{Methods}

\subsection{Stochastic cell rescaling}

We consider a system composed of $N$ atoms with coordinates and momenta
$\bm{q}_{i}$ and $\bm{p}_{i}$ contained in a box of volume $V$. For compactness,
in the following we indicate the vector of all coordinates and momenta
as $q$ and $p$, respectively. According to Hamilton equations, $p$
and $q$ evolve as
\begin{subequations}
\begin{align}
d\bm{p}_{i} & =-\frac{\partial U}{\partial\bm{q}_{i}}dt\label{eq:hami-p}\\
d\bm{q}_{i} & =\frac{\bm{p}_{i}}{m_{i}}dt\label{eq:hami-q}
\end{align}
\end{subequations}
where $t$ is the time, $m_{i}$ is the mass of the $i$-th particle, and $U$ the potential
energy of the system, which depends on the positions $q$.

To obtain ensemble averages in the $NPT$ ensemble, where the number of particles,
the external pressure, and the temperature are constant, the volume $V$
has to be allowed to fluctuate so as to sample states with probability
\begin{equation}
\mathcal{P}(p,q,V)\propto e^{-\frac{K+U+P_0V}{k_{B}T}}~.\label{eq:npt} 
\end{equation}
Here, $K=\sum_ip_i^2/(2m_i)$ is the kinetic energy of the system,
 $P_0$ is the external pressure, $k_B$ is the Boltzmann constant and $T$ is the temperature.
Equation~\ref{eq:npt} can be equivalently
written as 
\begin{equation}
-k_{B}T\log \mathcal{P}(p,q,V)=K+U+P_0V+C\label{eq:log-npt}
\end{equation}
where $C$ is an arbitrary constant.

The goal of a barostat is to induce changes in the volume $V$
that preserve the $NPT$ distribution. These changes are typically
made at constant scaled positions $s=q/\sqrt[3]{V}$, so that Cartesian positions
$q$ have to be uniformly scaled. Many algorithms
implement this change at constant scaled momenta $\pi=p\sqrt[3]{V}$ as well, thus
scaling Cartesian positions $q$ and momenta
$p$ with inverse factors.\cite{andersen1980molecular,parrinello1980crystal,parrinello1981polymorphic,nose1983constant,hoover1985canonical,melchionna1993hoover,sturgeon2000symplectic,kalibaeva2003constant,marry2007trotter,bussi2009isothermal,raiteri2011reactive,feller1995constant,kolb1999optimized,di2015stochastic,gao2016sampling}
This scaling preserves the phase-space volume
and originally stems from the use of a Hamiltonian formalism in the Andersen barostat.\cite{andersen1980molecular}
In other methods, this scaling factor is modified by a small correction that vanishes as the number of atoms in the simulated box grows.
\cite{martyna1994constant,zhang1995computer,martyna1996explicit,quigley2004langevin,yu2010measure,lippert2013accurate,cajahuaringa2018stochastic}
However, provided one evaluates the compression factor properly, scaling momenta is not strictly necessary, as suggested in Ref.~\onlinecite{gronbech2014constant}.
In the derivation below, we will assume that momenta are scaled with an inverse factor with respect to positions.
A similar algorithm can be derived assuming that momenta are not scaled (see Sec.~\ref{sec:scaling-moment}).

To obtain a continuous trajectory, we consider changes of $V$ obtained
by the solution of a first-order differential equation.
The most general first-order differential equation for the single variable $V$ with a preassigned
stationary distribution $\mathcal{P}(V)$ is stochastic and has the following form:
\begin{equation}
dV=D\frac{\partial\log (D\mathcal{P})}{\partial V}dt+\sqrt{2D}dW~.\label{eq:generic-sde}
\end{equation}
Here $dW$ is a Wiener noise and the equation is written
using Ito stochastic calculus.\cite{gardiner2009handbook} $D$ is an arbitrary function of the
volume $V$ that can be interpreted as a diffusion coefficient.
The stationarity of distribution $\mathcal{P}$ can be demonstrated
by considering the associated Fokker-Planck equation $\frac{\partial\mathcal{P}}{\partial t}=-\frac{\partial}{\partial V}\left(D\mathcal{P}\frac{\partial\log D\mathcal{P}}{\partial V}-\frac{1}{2}\frac{\partial}{\partial V}\left(2D\mathcal{P}\right)\right)=0$.
By inserting Eq.~\ref{eq:log-npt} in Eq.~\ref{eq:generic-sde}, with
simple manipulation, the most general stochastic differential equation for $V$ preserving the
isothermal-isobaric distribution can be shown to be in the form
\[
dV=-\frac{D}{k_{B}T}\left(\frac{\partial K}{\partial V}+\frac{\partial U}{\partial V}+P_0-\frac{k_{B}T}{D}\frac{\partial D}{\partial V}\right)dt+\sqrt{2D}dW~.
\]
We notice that the term $\frac{\partial U}{\partial V}$ is meant to be taken at fixed scaled coordinates
and corresponds
to the negative of the contribution of the potential energy to the internal pressure.
The term $\frac{\partial K}{\partial V}$ instead has to be computed at fixed scaled momenta,
thus defining $K=V^{-2/3}\sum_i\pi_i^2/(2m_i)$.
We thus have $\frac{\partial K}{\partial V}=-\frac{2K}{3V}$, and this term
corresponds to the negative of the kinetic contribution to the internal pressure.
The center-of-mass contribution to the internal pressure might be optionally included (see Supplementary Material, Sec.~I).

We then arbitrarily set $D=\frac{\beta_{T}Vk_{B}T}{\tau_{P}}$,
where $\beta_{T}$ is an estimate of the isothermal compressibility of the
system and $\tau_{P}$ is a characteristic time associated to the
barostat. The resulting differential equation is
\[
dV=-\frac{\beta_{T}V}{\tau_{P}}\left(\frac{\partial K}{\partial V} + \frac{\partial U}{\partial V}+P_0-\frac{k_{B}T}{V}\right)dt+\sqrt{\frac{2k_{B}T\beta_{T}V}{\tau_{P}}}dW~.
\]
This equation on $V$ can be converted to an equivalent equation on the strain $\epsilon=\log (V/V_0)$,
where $V_0$ is a reference volume that is needed to have the correct dimensionality,\cite{matta2011can}
and can be arbitrarily set to $V_0=1\textrm{nm}^3$.
By means of the Ito chain rule,\cite{gardiner2009handbook} one obtains
\begin{equation}
d\epsilon=-\frac{\beta_{T}}{\tau_{P}}\left(P_0-P_{\textrm{int}}\right)dt+\sqrt{\frac{2k_{B}T\beta_{T}}{V\tau_{P}}}dW\label{eq:de}~,
\end{equation}
where $P_{\textrm{int}}=\frac{2K}{3V}-\frac{\partial U}{\partial V}$ might be computed either including or
excluding the center-of-mass contribution (see Supplementary Material, Sec.~I).
The interpretation of Eq.~\ref{eq:de} is straighforward.
The deterministic term increases or decreases the volume when the internal pressure is
respectively larger or smaller than the external one.
Thanks to our definition of $D$, this term is equivalent to the one
used in the popular Berendsen barostat.\cite{berendsen1984molecular}
The stochastic term, however, allows volume fluctuations to be properly controlled.
Equation~\ref{eq:de} can be also derived by taking the high-friction limit
of a Langevin piston algorithm\cite{feller1995constant,bussi2009isothermal} with a volume-dependent friction,
provided the additional drift is taken into account correctly\cite{hottovy2015smoluchowski} (see Supplementary Material, Sec.~II).
Equation \ref{eq:de}
can then be coupled with Hamilton equations \ref{eq:hami-p} and \ref{eq:hami-q}
and with a thermostat
to sample the isothermal-isobaric ensemble. 
The increment in the volume logarithm leads to a rescaling of the cell matrix.
We thus refer to this scheme as stochastic cell rescaling.

\subsection{Working with unscaled momenta}
\label{sec:scaling-moment}

The derivation reported in the previous Section assumes
that volume changes are operated at constant scaled positions $s$ and momenta $\pi$.
It is possible to keep scaled positions $s$ and physical momenta $p$ fixed, instead, resulting
in an algorithm where momenta are not scaled when volume changes.
In this case, an extra factor $V^N$ has to be inserted in Eq.~\ref{eq:npt},
that in turn results in the use of the external temperature in evaluating the ideal gas pressure
contribution in $P_{\textrm{int}}$. At the same time, the gradient of the kinetic energy with respect to the
volume used in Eq.~\ref{eq:generic-sde} would be zero.
It can be seen that this change simply leads to the need to compute the internal pressure
using the average kinetic energy instead of the instantaneous one.\cite{gronbech2014constant}

In brief, two formulations of our algorithm are possible:
\begin{itemize}
\item
A formulation where momenta are scaled by a factor that is the inverse of the
factor used to scale the positions, and the internal pressure is calculated as
$P_{\textrm{int}}=\frac{2K}{3V} - \frac{\partial U}{\partial V}$.
\item
A formulation where momenta are untouched while scaling the simulation box, and
the internal pressure is calculated as 
$P_{\textrm{int}}=\frac{Nk_{B}T}{V} - \frac{\partial U}{\partial V}$.
\end{itemize}
If using a global thermostat so that center-of-mass momentum is conserved, $N$ should be replaced with $N-1$.
See Supplementary Material, Sec.~I, for further discussion on the center-of-mass contribution to the internal pressure.

We notice that most barostat implementations choose the first formulation, which
makes deriving a reversible integrator slightly more complex since positions and
momenta must be evolved simultaneously.
However, this formulation might be convenient
when using constraints\cite{ryckaert1977numerical,hess1997lincs}
or rigid bodies.\cite{miyamoto1992settle}
In these cases, indeed, the probability distributions of positions and momenta do not factorize,
making the ideal gas contribution more difficult to compute,
whereas the kinetic tensor implicitly contains the information about the position-dependent distribution of momenta.
Ref.~\onlinecite{gronbech2014constant} uses the second formulation instead.
Also the Monte Carlo barostat described in Refs.~\onlinecite{chow1995isothermal,aaqvist2004molecular} does not scale momenta, and indeed
includes explicitly a $(V'/V)^N$ term in the acceptance calculation, where $V'$ is the proposed volume.
The Berendsen barostat\cite{berendsen1984molecular} uses a hybrid formulation where the pressure is computed
using the instantaneous kinetic energy, as in the first formulation, but the momenta are not scaled when changing the simulation volume,
as in the second formulation.

The use of the instantaneous kinetic energy instead of its average value
can be seen as a source of noise in the dynamics of the volume. In particular, the kinetic energy has
fluctuations equal to $\sqrt{\frac{3N}{2}}k_BT$ and an autocorrelation time $\tau_K$
that is system dependent and that can be controlled by choosing the parameters
of the thermostat.
These fluctuations can be approximated as an additional noise term in Eq.~\ref{eq:de}.
If $\tau_K$ is significantly smaller than the relaxation time of the volume ($\tau_K\ll\tau_P$)
this noise can be considered as white and equal to
$\frac{\beta_T}{\tau_P}\frac{2}{3V} \sqrt{\frac{3N}{2}}k_BT \sqrt{\tau_K}dW
=
\frac{\beta_Tk_BT}{\tau_PV}\sqrt{\frac{2N\tau_K}{3}} dW
$.
The ratio between the standard deviation of this noise and the standard deviation of the
random noise in Eq.~\ref{eq:de} is
\begin{equation}
\frac{\frac{\beta_Tk_BT}{\tau_PV}\sqrt{\frac{2N\tau_K}{3}}}{\sqrt{\frac{2k_{B}T\beta_{T}}{V\tau_{P}}}}=
\sqrt{\frac{\tau_K Nk_BT \beta_T}{3\tau_PV}}\approx
\sqrt{\frac{\tau_K N\sigma_V^2}{3\tau_P\langle V\rangle^2}}~.
\label{eq:noise-ratio}
\end{equation}
In the last step we exploited the fact that $\beta_T=\frac{\sigma^2_V}{\langle V \rangle k_BT}$, where $\sigma_V^2$ are
the fluctuations of the volume, and we approximated $V$ with its average value $\langle V \rangle$.
Since in condensed phases we expect the relative volume fluctuations to be smaller than $\sqrt{1/N}$,
and we assumed $\tau_K\ll\tau_P$, the contribution of the additional noise
is negligible. This is the case for most of the practical applications considered here,
with the used thermostat settings.
However,
in general settings, the two sources of noise might be comparable.
It is thus important to implement stochastic cell rescaling so that,
if the internal pressure is computed using the instantanous kinetic energy,
momenta are scaled with the correct factor whenever the volume changes.

\subsection{Effective energy drift}
\label{sec:conserved-energy}

Stochastic cell rescaling is based on the use of Hamilton equations \ref{eq:hami-p} and \ref{eq:hami-q} and of
the stochastic differential equation \ref{eq:de}, both satisfying detailed balance.
When they are integrated with a finite time step algorithm, however, detailed balance is violated.
This violation can be monitored during the simulation\cite{bussi2007canonical,bussi2007accurate}
and used to determine if the time step and
the other simulation parameters were chosen correctly, to verify that forces are correctly computed as
the negative derivatives of the energy function,
or to compute the acceptance for a so-called Metropolized integrator.\cite{scemama2006efficient}
This contribution can be interpreted as the work performed by the integration algorithm on the system.\cite{sivak2013using}
In the case of pure Hamilton equations, this drift corresponds exactly to the change in the total energy
of the system. When a thermostat is used, its contribution to the drift has to be added.\cite{bussi2007canonical,bussi2007accurate}
Similarly, the barostat will contribute to the drift.

In order to compute the contribution of the barostat to the drift, it is necessary
to compute the relative probabilities of generating forward steps, where $\epsilon\rightarrow\epsilon'$,
and backward steps, where $\epsilon'\rightarrow\epsilon$.
We notice that the prefactor of the stochastic term in Eq.~\ref{eq:de} depends explicitly on $\epsilon$.
As discussed in more details in Supplementary Material, Sec.~III, this dependence might increase detailed-balance violations.
By performing a change of variable to $\lambda=e^{\epsilon/2}\sqrt{V_0}=\sqrt{V}$, instead,
by means of the Ito chain rule,\cite{gardiner2009handbook}
one obtains the following differential equation
\begin{equation}
d\lambda=-\frac{\beta_{T}\lambda}{2\tau_{P}}\left(P_0-P_{\textrm{int}}-\frac{k_BT}{2V}\right)dt+\sqrt{\frac{k_{B}T\beta_{T}}{2\tau_{P}}}dW~.\label{eq:dlambda}
\end{equation}
In this equation, the prefactor of the stochastic term is a constant.
By simply integrating this equation with finite difference increments, the drift can be computed in the same way as it is
computed in the high-friction limit of the Langevin equation,\cite{bussi2007accurate} and corresponds to the calculation
of the acceptance in the so-called smart Monte Carlo method\cite{rossky1978brownian} (see Supplementary Material, Sec.~IV).

It it important to recall that
the effective energy drift quantifies how much detailed balance is violated. However, detailed balance is not
a necessary condition for reaching the target stationary distribution.\cite{manousiouthakis1999strict}
In addition, it has been shown that this drift might overestimate the errors observed
in sampling the configurational degrees of freedom.\cite{fass2018quantifying}

\subsection{Integration of the equations of motion}

We considered three different ways to integrate Eqs.~\ref{eq:hami-p}, \ref{eq:hami-q}, and \ref{eq:de}.
In all of them,
it is possible to postpone the calculation of the virial to every $N_P$ steps
in a multiple-time-step fashion\cite{tuckerman1990molecular}
in order to speed up the calculation.
In the last two algorithms, it is possible to define the effective energy drift (see Sec.~\ref{sec:conserved-energy}).
More details are reported in Supplementary Material, Sec.~V.

\emph{Euler integrator}.
The volume is evolved by propagating its logarithm
using a finite time step approximation of Eq.~\ref{eq:de}.
Positions and velocities are then evolved using velocity Verlet.
Forces \emph{are not} recomputed after the volume change. As a consequence,
the obtained trajectory \emph{is not} reversible.

\emph{Reversible Euler integrator}. The volume is evolved  by propagating its square root
using a finite time step approximation of Eq.~\ref{eq:dlambda}.
Positions and velocities are then evolved using velocity Verlet.
Forces \emph{are} recomputed after the volume change. As a consequence,
the obtained trajectory \emph{is} reversible.
However, this is paid with an extra force calculation every $N_P$ steps.
In order to quantify the effective energy drift, the virial needs to be recomputed after the volume change.

\emph{Trotter-based integrator}. The volume is evolved  by propagating its square root
using a finite time step approximation of Eq.~\ref{eq:dlambda}.
Positions and velocities are evolved using velocity Verlet simultaneously with volume change using a Trotter splitting.
There is no need to recompute forces after the volume change.
However, in order to quantify the effective energy drift, the virial needs to be computed also
at the step immediately after the one at which scaling was applied.
The obtained trajectory \emph{is} reversible.

For simplicity, we decided to use Eq.~\ref{eq:de} in the Euler integrator, where
the effective energy drift would not be well-defined anyway, whereas we used Eq.~\ref{eq:dlambda}
for the reversible implementations.
All the three schemes can be implemented either scaling or not scaling velocities upon
volume change (see Sec.~\ref{sec:scaling-moment}).
For each of the integrators, Table \ref{tab-int} summarizes the cost in term of how many force and virial
calculations are required on average for each simulation step.
These integrators,
similarly to those discussed in Ref.~\onlinecite{lippert2013accurate},
can in principle be used with an arbitrarily large $N_P$, provided that $\tau_P$
is also chosen large enough.

\begin{table}
\centering
 \begin{tabular}{|c c c c|} 
 \hline
Integrator & Reversible & $N_{\textrm{forces}}$  & $N_{\textrm{virial}}$  \\
 \hline
Euler             & no         & $1$                      & $1/N_P$                    \\
Reversible Euler  & yes        & $1+1/N_P$                & $2/N_P$                  \\
Trotter           & yes        & $1$                      & $\min\left(1,2/N_P\right)$ \\
 \hline
 \end{tabular}
\caption{
\label{tab-int}
Computational overhead for the discussed integrators.
Integrators are named as discussed in the main text.
$N_P$ is stride for the propagation of the barostat.
$N_{\textrm{forces}}$ and $N_{\textrm{virial}}$
are the average number of times forces and virial need to be calculated for every MD step, respectively.
For all the integrators, the equation of the barostat is propagated every $1/N_P$ steps.
}
\end{table}

\subsection{Semi-isotropic version}

Eq.~\ref{eq:de} can be generalized to cases where $\epsilon$ is a matrix representing the
deformation of the system. We here derive the equations required to sample the constant
surface-tension ensemble $NP\gamma T$, where the equilibrium probability reads\cite{zhang1995computer}
\begin{equation}
\mathcal{P}(p,q,A,L)\propto e^{-\frac{K+U+P_0AL-\gamma_0A}{k_{B}T}}~.
\label{eq:p-npgt}
\end{equation}
Here $A$ is the area of the simulation box in the $xy$ plane, $L$ is its height, and $\gamma_0$ is the surface tension multiplied by the number of surfaces.

We arbitrarily set the diffusion coefficients for $A$ and $L$ as $D_{A}=\frac{2\beta_TA^2k_BT}{3V\tau_P}$ and $D_L=\frac{\beta_TL^2k_BT}{3V\tau_P}$,
respectively.
By defining the variables $\epsilon_{xy}=\log (A/A_0)$ and $\epsilon_z=\log (L/L_0)$,
 and following a procedure similar to the one above, these equations of motion are obtained:
\begin{subequations}
\begin{equation}
d\epsilon_{xy}=-\frac{2\beta_T}{3\tau_P}\left(
P_0 - \frac{\gamma_0}{L} - \frac{P_{\textrm{int},xx}+P_{\textrm{int},yy}}{2}\right)dt
+\sqrt{\frac{4k_BT\beta_T}{3V\tau_P}}dW_{xy}
\label{eq:dexyz-xy}
\end{equation}
\begin{equation}
d\epsilon_z=-\frac{\beta_T}{3\tau_P}\left(
P_0 - P_{\textrm{int},zz}\right)dt
+\sqrt{\frac{2k_BT\beta_T}{3V\tau_P}}dW_{z}
\label{eq:dexyz-z}
\end{equation}
\label{eq:dexyz}
\end{subequations}
Here $P_{\textrm{int},xx}$, $P_{\textrm{int},yy}$, and $P_{\textrm{int},zz}$ are components of the internal pressure tensor
and the two noise terms $dW_{xy}$ and $dW_{z}$ are explicitly written with different subscripts to remark that
they have to be drawn independently. For an extensive derivation of Eqs.~\ref{eq:dexyz} see Supplementary Material, Sec.~VI.

By taking the sum of the two equations above and setting $\gamma_0=0$, one obtains Eq.~\ref{eq:de}.
This means that, if no external tension is applied, the dynamics of the volume in the semi-isotropic case will be identical
to the isotropic case.
In principle, it is possible to tune separately the compressibility
of the system in the $xy$ and $z$ directions (see Supplementary Material, Sec.~VII, for the special case where
cell height $L$ is kept constant), or even to choose a non-diagonal diffusion matrix for the $A$
and $L$ variables. This choice would only affect the timescale at which $A$ and $L$ equilibrate, leaving
the sampled distribution unchanged.

\subsection{Computational details}

Simulations of the Lennard-Jones fluid were performed using a modified version of the SimpleMD program.
A system of $N=256$ particles was simulated in a cubic box with a time step of 0.005
for $10^7$ steps, accumulating statistics every 10 steps. Forces were truncated at distance 2.5.
Temperature was set to $T=1.5$ and controlled with a stochastic velocity rescaling
thermostat\cite{bussi2007canonical} with relaxation time $\tau_T=0.05$.
Pressure was set to $P_0=1$ and controlled using stochastic cell rescaling
with a range of control parameters.
All parameters are reported in reduced Lennard-Jones units.
All the reported quantities were computed discarding the initial $2.5\times10^6$ steps.

Simulations of the liquid water, host-guest and guest only, and the membrane systems were performed with a modified version of GROMACS 2019.4.\cite{abraham2015gromacs}
The liquid water system comprised 2850 waters in a rhombic dodecahedron box. 
The TIP3P model\cite{jorgensen1983comparison} was used to represent the water molecules. A short equilibration run lasting 500 ps was first conducted in the $NVT$ ensemble. 
The production phase consisted of three sets of $NPT$ simulations using the Parrinello-Rahman,\cite{parrinello1981polymorphic} Berendsen,\cite{berendsen1984molecular} and stochastic cell rescaling barostats, respectively. 
In each set, a range of control parameters was explored, using 1 bar as reference for isotropic pressure coupling in all cases. All simulations lasted 10 ns 
and statistics were saved every 200 steps (0.4 ps). The reference temperature was 300 K in all runs and was controlled through a stochastic velocity rescaling 
thermostat \cite{bussi2007canonical} with a relaxation time $\tau_T=0.1$ ps. A Verlet cut-off scheme was employed for neighbor searching, updating the neighbor list every 10 steps.  
All shown results were obtained discarding the first 2 ns of simulation.

A smaller water box, comprising 900 TIP3P water molecules, was employed to perform the physical validation tests indicated 
in Refs.~\onlinecite{shirts2013simple,merz2018testing}. To conduct these tests, the physical\_validation package, an open-source and platform-independent Python library 
(https://physical-validation.readthedocs.io) in which they are implemented, was used. 
For each barostat (Berendsen, Parrinello-Rahman and stochastic cell rescaling), 
two simulations of 10 ns each were run at the reference pressure of 1 bar and 301 bar, respectively, 
with $N_P=10$ steps and $\tau_P =1$ ps for Berendsen and Parrinello-Rahman and $\tau_P =0.5$ ps for stochastic cell rescaling.
All GROMACS input files were taken and adapted from the examples/water\_ensemble subfolder coming with the package.

Free energy differences for the host-guest (OA-G3) and guest only (G3) systems were computed from expanded ensemble simulations in solution, conducted using coordinates, topology, and set-up provided by Ref.~\onlinecite{rizzi2020sampl6}. 
A detailed description of the simulation parameters can be found therein. Briefly, decoupling of the guest was achieved by completely turning off charges first and then removing Van der Waals interactions in both the host-guest and guest only systems. 
The entire procedure comprised a total of 40 lambda windows. A velocity Verlet integrator was used with a time step of 2 fs. 
The expanded ensemble simulations were divided in two stages: an initial equilibration stage to adaptively estimate the expanded ensemble weights and 
the following production phase in which the weights were kept fixed. For the host-guest system, 60 ns were required for the equilibration of weights 
with Berendsen and stochastic cell rescaling, while for the guest only systems about 30 ns were necessary in all cases. 
From the production stage, free-energy differences were computed with the multistate Bennett acceptance ratio (MBAR) method\cite{shirts2008statistically}
using the alchemical\_analysis tool version 1.0.2\cite{Klimovich:2015er} along with pymbar version 3.0.5.
The free energy associated to the presence of the restraints on the center of mass in the decoupled state was computed as
$\Delta G_{\textrm{restr}}=-\frac{3}{2}\log \left(\frac{2\pi k_BT}{k}\right) + \log V_{\textrm{mol}}=16.73$ kJ/mol,
where $k=1000$ kJ/(mol$\cdot$nm$^2$) is the coupling constant and $V_{\textrm{mol}}=1.66$ nm$^3$ is the volume corresponding to the one molar standard state.

Replica exchange simulations for decoupling of the guest only in solution were performed using 40 replicas corresponding to the windows employed to turn off 
charges and removing Van der Waals interactions in the expanded ensemble simulations, thus resulting in a Hamiltonian replica-exchange protocol. 
Two variants of the system with varying size of the box were considered: the same used for the expanded ensemble simulations, taken from Ref.~\onlinecite{rizzi2020sampl6},
where a distance of 1.2 nm from all guest heavy atoms in all directions was applied, and a smaller one where such distance was set to 0.8 nm. 
A leap-frog integrator was used with a time steps of 2 fs and exchanges between replicas were attempted every 400 steps (0.8 ps). 
Production runs were conducted for a total of 5 ns/replica and were used to compute the free-energy differences through 
Bennett’s acceptance ratio method (BAR)\cite{bennett1976efficient} as implemented in the gmx bar module of GROMACS. 
Decoupling of the guest E20 (donepezil, extracted from PDB ID: 1EVE) was conducted according to the same set-up. 
Ligand parameters were determined using the General Amber Force Field (GAFF)\cite{wang2004development} following the RESP procedure\cite{bayly1993well}
to determine the molecule charges. During the simulations, the conformation of the ligand was kept fixed by restraining the 
root-mean-square displacement from the crystal conformation computed after superimposing the two structures.
To this end, the RESTRAINT feature of the bias module of PLUMED\cite{tribello2014plumed} was used, setting a force constant of 209200 kJ/(mol$\cdot$nm$^2$)
for the harmonic restraint.

The membrane system was built as described in the ``KALP$_{15}$ in DPPC'' (the KALP model peptide in a lipid bilayer of dipalmitoylphosphatidylcholine) 
tutorial,\cite{ lemkul2018proteins} which protocol was based on a previous work\cite{kandasamy2006molecular} and 
used the GROMOS96 53A6 force field,\cite{oostenbrink2004biomolecular} extended to include Berger lipid parameters.\cite{ berger1997molecular} 
The procedure included usage of the InflateGRO methodology\cite{ kandt2007setting} to pack the lipids around the 
embedded protein. After solvation, a system comprising 126 DPPC lipid and 4182 SPC water molecules\cite{berendsen1981interaction} was obtained.
Energy minimization and equilibration of the system were conducted as described in detail in the tutorial. 
As for the liquid-water system, the production phase consisted of three sets of $NPT$ simulations 
using the Parrinello-Rahman, Berendsen and stochastic cell rescaling barostats.
A reference pressure of 1 bar was employed for semi-isotropic 
pressure coupling in all cases, along with $\tau_P =2$ ps and a compressibility of 4.5x10$^{-5}$ bar$^{-1}$. The temperature 
was set to 323 K and was controlled through the Nose-Hoover thermostat\cite{nose1984unified,hoover1985canonical} with a relaxation time $\tau_T=0.5$ ps, 
coupling separately a first group comprising the protein and DPPC lipids and a second one including the solvent and ion molecules. All simulations 
lasted 100 ns and statistics were saved every 200 steps (0.4 ps). 
All shown results were obtained analyzing the second half of the trajectory.

In all simulations, statistical errors were determined using block analysis \cite{flyvbjerg1989error}  with a variable number of blocks and conservatively using
the largest estimate for the error.
For all the simulations performed using the Parrinello-Rahman barostat, fluctuations were imposed to be isotropic (Lennard-Jones fluid, liquid water and guest-only systems)
or semi-isotropic (membrane system), as implemented in GROMACS.
Indeed, the fully flexible version of the Parrinello-Rahman barostat is known to be unstable in these cases.
The resulting equations of motion for the box are thus not exactly equivalent to those
of the Andersen\cite{andersen1980molecular} or Martyna-Tobias-Klein\cite{martyna1994constant} algorithms, but are representative
of a second-order barostat.

\section{Numerical tests}

\subsection{Lennard-Jones fluid}

\begin{figure}
\includegraphics[width=\columnwidth]{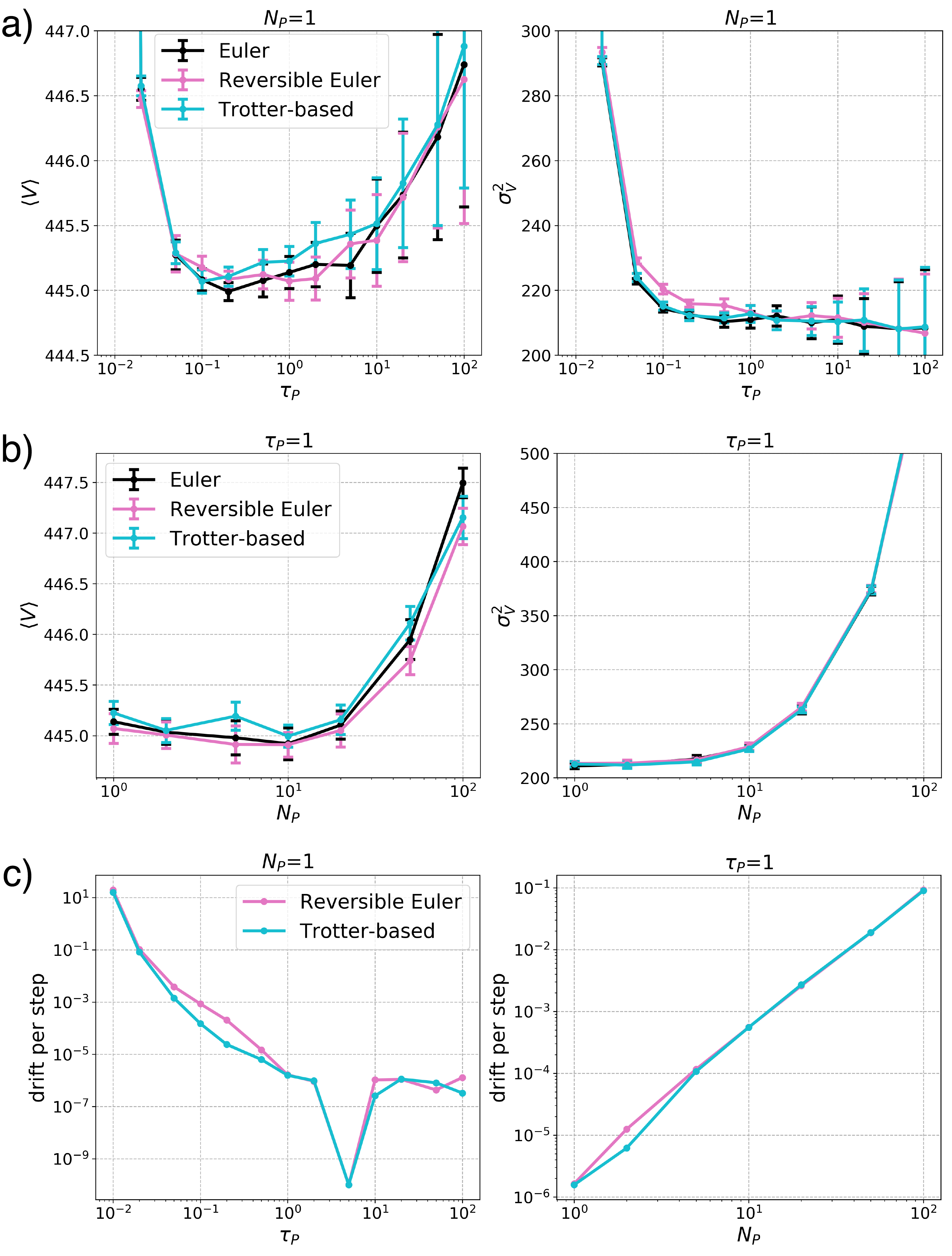}
\caption{
Results from simulations of a Lennard-Jones fluid.
a) Average and fluctuations of the volume (left and right panels, respectively) as a function of the time constant for pressure coupling ($\tau_P$)
at fixed frequency for pressure coupling ($N_P=1$).
b) Average and fluctuations of the volume (left and right panels, respectively) as a function of the frequency for pressure coupling ($N_P$)
at fixed time constant for pressure coupling ($\tau_P=1$).
c) Effective energy drift per step, obtained from the slope of a line interpolating the effective energy drift
on the entire trajectory.
With some settings, a negative slope is obtained and is here shown as $10^{-10}$.
\label{fig_lj}
}
\end{figure}

We tested stochastic cell rescaling on the simulation of a Lennard-Jones fluid in the $NPT$ ensemble.
We first performed a range of simulations at constant
volume, computed the average internal pressure, and integrated it so as to obtain a reference distribution
for the volume $V$. The predicted volume fluctuations correspond to an isothermal compressibility $\beta_T\approx0.3$.
This value is used as an input in the simulations performed with the barostat. The distribution of the volume obtained using
barostat parameters $N_P=1$ and $\tau_P=1$ is perfectly overlapping
with the reference one (see Fig.~S1).

We then evaluated the robustness of the results by monitoring volume average and fluctuations when changing
barostat parameters, testing the three discussed integrators (Fig.~\ref{fig_lj}).
We first fixed $N_P=1$ and investigated the dependence of the results on $\tau_P$ (Fig.~\ref{fig_lj}a).
Any $\tau_P\ge 0.1$ report results consistent with the reference. As expected,
the statistical error on the volume and on its fluctuations grows with $\tau_P$. This suggests
that, in order to equilibrate and sample the volume variable as quickly as possible, $\tau_P$ should
be chosen as small as possible. The autocorrelation time of $V$ can be seen to be
close to $\tau_P$ when $\tau_P$ is large enough (see Fig.~S2).
The relationship between stochastic cell rescaling and the Langevin piston approach
can be appreciated by comparing the autocorrelation function of the volume in a Langevin piston
with decreasing values of the barostat mass (see Fig.~S3).
We then fixed $\tau_P=1$ and investigated the dependence of the results on $N_P$ (Fig.~\ref{fig_lj}b).
$N_P\ge 10$ resulted in a volume variance observably larger than its reference value.
Interestingly, the three introduced integrators resulted in very similar accuracy when used with the same parameters.

For the two integrators that allow an effective energy drift to be defined, we computed this drift for all the chosen
sets of parameters (Fig.~\ref{fig_lj}c). As long as $\tau_P\ge 1$ and $N_P=1$ the drift was of the order of
$10^{-6}$ energy units per step, comparable to the one obtained in $NVT$ simulations.
By testing different values of $N_P$, we observed that the drift steadily grew with $N_P$.
This drift can be used to estimate if the violations of detailed balance induced by the barostat
are exceeding those that are present also in absence of the barostat.\cite{bussi2009isothermal}

All the reported results were obtained using the formulation in which momenta are scaled
when volume changes (see Sec.~\ref{sec:scaling-moment}). Results obtained without scaling momenta
were equivalent and are reported in Fig.~S4.

\subsection{Liquid water}

\begin{figure}
\includegraphics[width=\columnwidth]{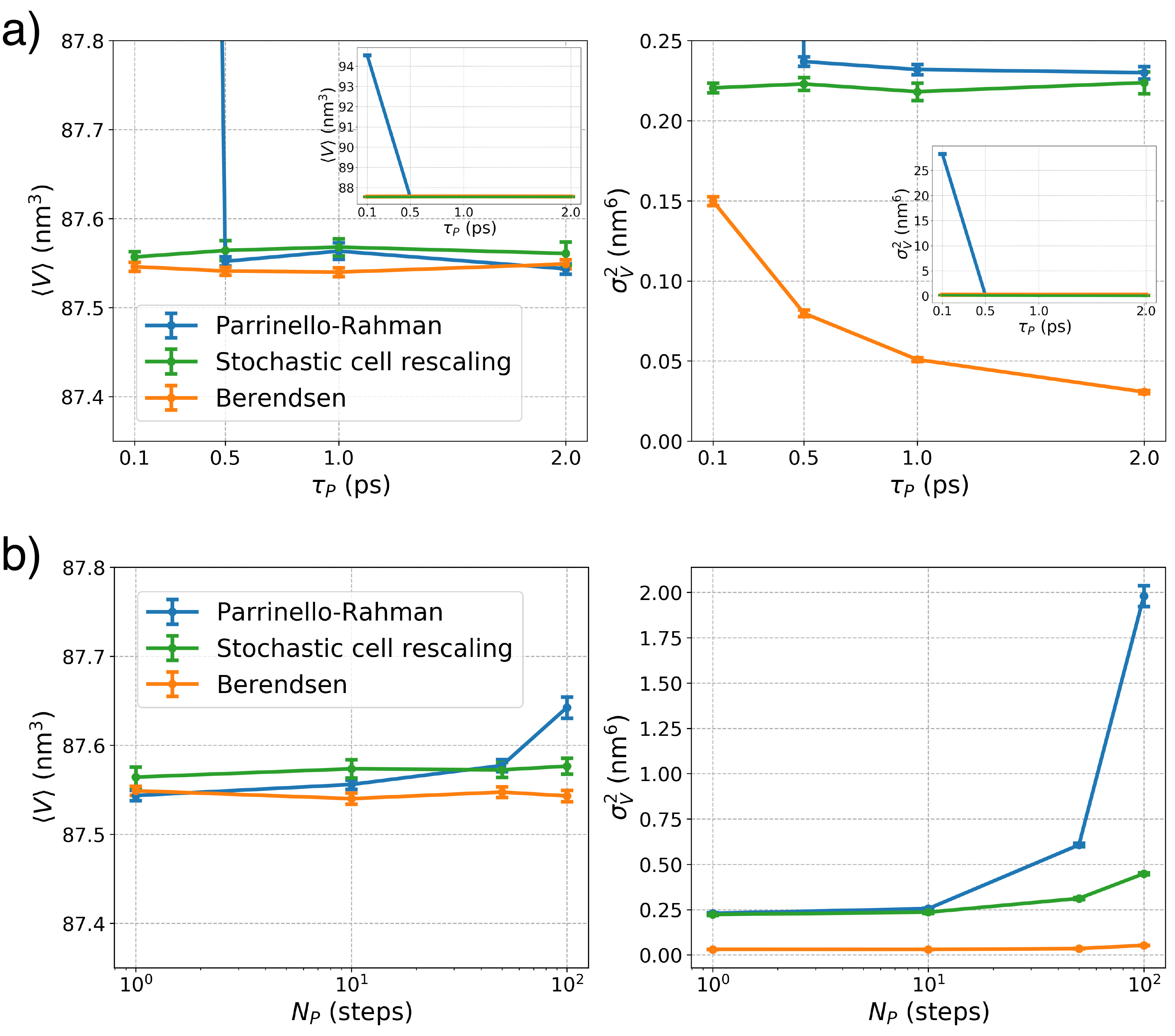}
\caption{
Results from simulations of a TIP3P water box.
a) Average and fluctuations of the volume (left and right panels, respectively) as a function of the time constant for pressure coupling ($\tau_P$)
at fixed frequency for pressure coupling ($N_P=1$).
b) Average and fluctuations of the volume (left and right panels, respectively) as a function of the frequency for pressure coupling ($N_P$)
at fixed time constant for pressure coupling ($\tau_P=2$ for Parrinello-Rahman and Berendsen, $\tau_P=0.5$ for stochastic cell rescaling).
\label{fig_water}
}
\end{figure}

We then tested stochastic cell rescaling on the simulation of a box of TIP3P water molecules.
These simulations were done using GROMACS, which includes an implementation of the Parrinello-Rahman
algorithm \cite{parrinello1981polymorphic} 
that is known to report results in the $NPT$ ensemble and can thus be used as a reference.
We also tested the Berendsen algorithm,\cite{berendsen1984molecular}
that neglects the noise term and is expected produce incorrect volume fluctuations.
In the comparison between the Parrinello-Rahman and the two other schemes, one should
consider that there is not an equivalence in the definition of $\tau_P$.
Thus, results at the same $\tau_P$ cannot be directed compared.

We first analyzed the dependence of volume averages and fluctuations as a function of $\tau_P$,
setting $N_P=1$ (Fig.~\ref{fig_water}a). Volume averages were always consistent in the three schemes,
but the Parrinello-Rahman implementation showed some instability for the smallest choice of $\tau_P=0.1$ ps.
Volume fluctuations obtained with Parrinello-Rahman and with stochastic cell rescaling were
consistent and lead to an estimate of the isothermal compressibility equal to
$\beta_T=\frac{\sigma_V^2}{\langle V\rangle k_BT}\approx 1.05\times 10^{-3}\frac{\textrm{nm}^3}{\textrm{kJ/mol}}\approx 6.3\times 10^{-5} \textrm{bar}^{-1}$.
These results are consistent with those reported in Ref.~\onlinecite{jorgensen1998temperature} for the same water model.
We notice that this value is markedly different from the experimental value $\beta_{T,\textrm{exp}}=4.5\times 10^{-5} \textrm{bar}^{-1}$.
This discrepancy is known for the TIP3P model.
The experimental compressibility has been here used as an input parameter.
The effect of choosing an input compressibility that is inconsistent with the compressibility of the simulated
system is only to change the effective relaxation time of the volume, as it can be seen by computing
its autocorrelation function (see Fig.~S5).
Fluctuations were significantly underestimated by the Berendsen barostat.
Counterintuitively, the fluctuations reported by the Berendsen barostat increased when $\tau_P$
decreased. We interpret this effect as a consequence of the fact that the amplitude of
the fictitious noise that is implicitly included in the Berendsen barostat by the usage 
of the instantaneous kinetic energy grows when $\tau_P$ decreases (see Eq.~\ref{eq:noise-ratio}),
partly compensating for the lack of an explicit noise term.

\begin{figure}
\includegraphics[width=\columnwidth]{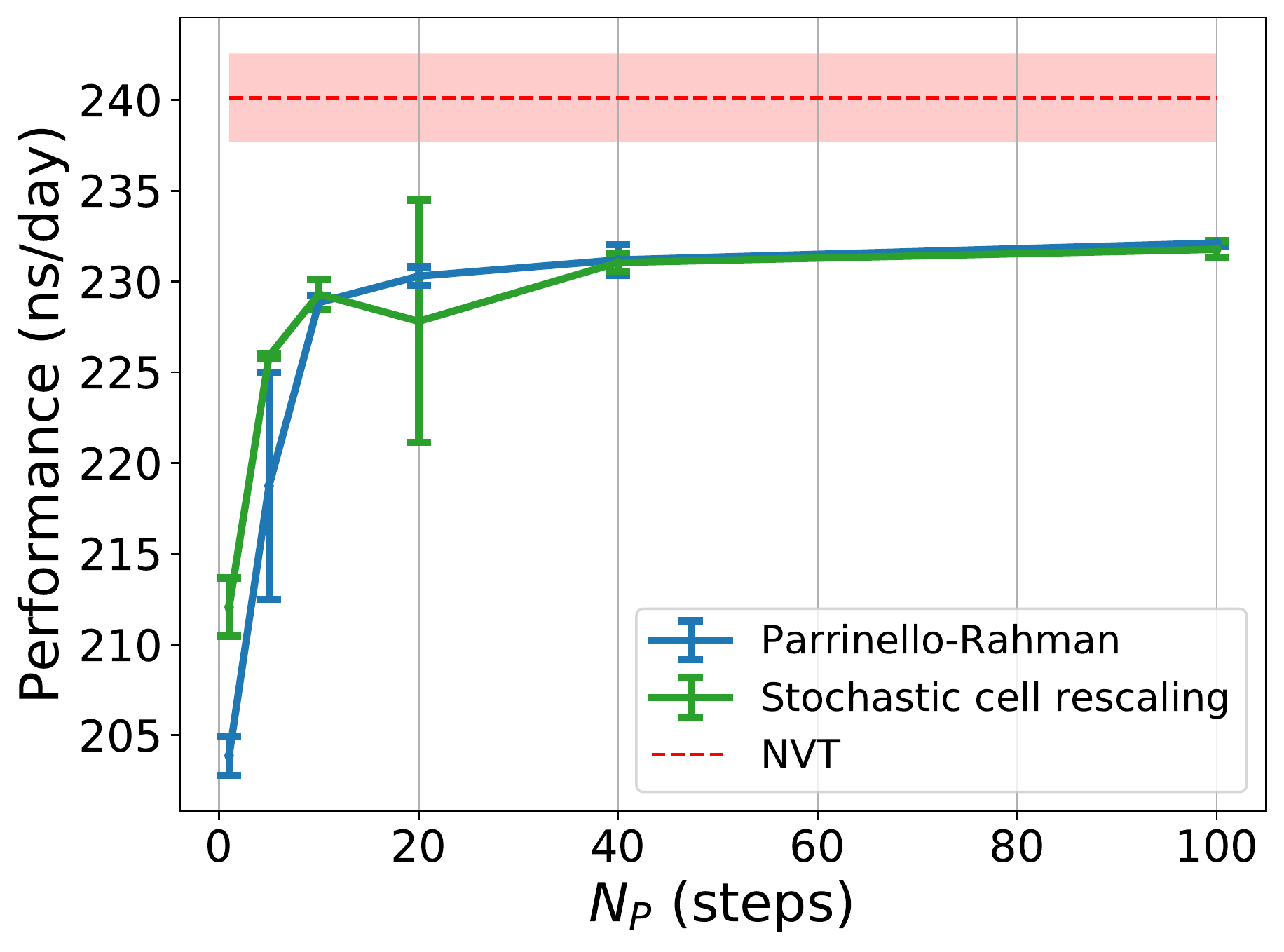}
\caption{
Dependence of performance on choice of $N_P$. Results are for the water box simulation in GROMACS
	and were produced using 4 cores on a Intel\textsuperscript{\textregistered} Xeon\textsuperscript{\textregistered} CPU E5-2620 v2 (2.10GHz) and 1 NVIDIA GeForce GTX TITAN GPU. Error bars report the standard deviation
over 5 simulations lasting 1 ns. The red dashed line and shade report the performance and standard deviation, respectively, of an $NVT$ simulation for the same system.
\label{fig_perform}
}
\end{figure}

We then analyzed the dependence of volume averages and fluctuations as a function of $N_P$,
setting $\tau_P=2$ ps for Parrinello-Rahman and Berendsen and $\tau_P=0.5$ 
ps for stochastic cell rescaling (Fig.~\ref{fig_water}b).
In all cases, a too large $N_P$ resulted in an overestimation of the fluctuations.
For this specific system and simulation parameters, $N_P=10$ seems a reasonable compromise for all barostats.
We notice that $N_P=10$ is already capable to give a significant performance boost (see Fig.~\ref{fig_perform}).

A validation similar to the one reported in Refs.~\onlinecite{shirts2013simple,merz2018testing} was then performed on a smaller water box.
The results are reported in Fig.~S6 and in Table S1, and confirm that
the quality of the volume distributions obtained with
our implementation of the stochastic cell rescaling algorithm is significantly more reliable than the one obtained
with the Berendsen barostat. Interestingly, stochastic cell rescaling appears as slightly closer
to the reference result than the Parrinello-Rahman barostat (Table S1),
although this might depend on the technical details of the integration scheme implemented in GROMACS.

\subsection{Free energy differences}

We then tested the impact of the barostat algorithm on the calculation of the solvation and absolute binding free energies for small molecules.
This is a relevant case since the incorrect volume fluctuations produced by
the Berendsen barostat were recently suggested to significantly affect absolute binding affinities.\cite{rizzi2020sampl6}
We here used the same settings reported in Ref.~\onlinecite{rizzi2020sampl6}, and in particular the same extended ensemble protocol,
on a host-guest system (OA-G3) and on a guest-only system (G3).
Since the extended ensemble protocol is not compatible with the Parrinello-Rahman implementation present in GROMACS, we only report
results obtained with stochastic cell rescaling and Berendsen algorithms.
The solvation free energies obtained with different barostats are
close to each other (see Table~\ref{table-fe_ee}). Their differences, which enter in the calculation of the absolute binding free energy,
agree within the respective statistical error.
We remark that the absolute binding free energies reported in Table~\ref{table-fe_ee} did not include corrections for
finite size effects, and should not be compared quantitatively with those reported in Ref.~\onlinecite{rizzi2020sampl6}.
However, these corrections are expected to be independent of the choice of the barostat.

In order to better investigate the possible effects of incorrect volume fluctuations on the calculation of solvation free energies,
we tested the guest-only system in a replica-exchange protocol
that is compatible with the Parrinello-Rahman barostat implemented in GROMACS.
In this case, the size of the water box was decreased so as to amplify the impact of the barostat algorithm (see Table~\ref{table-fe_solvation}).
For all the tested settings, the discrepancy between the results obtained with the Berendsen, stochastic cell rescaling, and 
Parrinello-Rahman barostats were below a fraction of kJ/mol. We notice that the reported estimate of the statistical error is likely
an underestimation since it does not take into account the exchange of coordinates between the replicas.
Tests performed with a larger ligand (E20), representative of typical drug molecules pursued in medicinal chemistry frameworks,
confirm that differences are small ($<1$ kJ/mol) and likely not correlated
or weakly correlated with the choice of the barostat.

\begin{table}
\centering
	\caption{Free-energy differences (kJ/mol) for decoupling of the guest 5-hexenoic acid in host-guest (OA-G3) and guest only (G3) simulations in solution performed with the Expanded Ensemble method.  
$\Delta G$ in the standard state is computed subtracting the contribution of the restraint acting on the guest in the decoupled state.
\label{table-fe_ee}
}
\begin{tabular}{ |c | c  c c | }
\hline
	& \textbf{OA-G3} & \textbf{G3} & $\Delta G$ \\
\hline
	Berendsen                  &  433.1 $\pm$ 0.6 & 389.0 $\pm$ 0.5 & -27.3 $\pm$ 0.7 \\
	Stochastic cell rescaling  &  432.6 $\pm$ 0.6 & 387.9 $\pm$ 0.5 & -28.0 $\pm$ 0.7 \\
\hline
\end{tabular}
\end{table}

\begin{table}
\centering
	\caption{Free-energy of solvation (kJ/mol) computed from decoupling simulations of the guest (G3) and ligand E20 in solution performed with the Hamiltonian Replica Exchange method. 
	The simulations of G3 were performed at varying box sizes: a larger one defined with a 1.2 nm distance (-d 1.2) from the guest in all directions,
	and a smaller one with a 0.8 nm distance (-d 0.8).
\label{table-fe_solvation}
}
\begin{tabular}{ |c|c c c|}
\hline
	& \textbf{G3 -d 1.2} & \textbf{G3 -d 0.8} & \textbf{E20}\\
\hline
	Berendsen  &  389.40 $\pm$ 0.05 & 389.65 $\pm$ 0.10 & 173.02 $\pm$ 0.14  \\
	Parrinello-Rahman  &  389.35 $\pm$ 0.17 & 389.03 $\pm$ 0.12 & 172.28 $\pm$ 0.13  \\
	Stochastic cell rescaling  &  389.41 $\pm$ 0.06 & 389.31 $\pm$ 0.14 & 172.70 $\pm$ 0.13 \\
\hline
\end{tabular}
\end{table}

\subsection{Membrane simulation}

We finally tested the impact of the barostat algorithm on the equilibration of a model membrane, including a short transmembrane protein.
In this case we employed the anisotropic version of the stochastic cell rescaling, Berendsen, and Parrinello-Rahman algorithms,
where no external tension is applied to the membrane. Whereas the simulations are probably too short to properly equilibrate this system,
it is clear that stochastic cell rescaling and Parrinello-Rahman provide consistent results,
whereas the Berendsen barostat leads to a suppression of the fluctuations both in the membrane surface and in the cell size in the
direction orthogonal to the membrane (Table \ref{table-membrane}).
Time series for the simulated trajectories are reported in Fig.~S7.

\begin{table*}
\centering
	\caption{Average ($\langle V\rangle$) and fluctuations (var($V$)) of the volume, area of the simulation box along the xy plane ($A$) and length of the box along the z axis ($L$) for a lipid membrane system. The system comprises a bilayer made up of DPPC lipids and the short transmembrane protein KALP$_{15}$.
	Note that the membrane is oriented in such a way that the lipid bilayer lies on the xy plane and the normal of the bilayer is aligned with the box z axis.
\label{table-membrane}
}
\begin{tabular}{ P{4cm}|P{4cm}| P{4cm} |P{4cm} }
        & \textbf{Parrinello-Rahman} & \textbf{Stochastic cell rescaling} & \textbf{Berendsen} \\
\hline
	$\langle V\rangle$ (nm$^3$)  & 280.30 $\pm$ 0.03 	& 280.23 $\pm$ 0.02 & 280.25 $\pm$ 0.03\\
\hline
	$\sigma_V^2$  (nm$^6$) & 0.723 $\pm$ 0.005 & 0.728 $\pm$ 0.010 & 0.227 $\pm$ 0.004 \\
\hline
	$\langle A\rangle$  (nm$^2$) & 38.03 $\pm$ 0.07 &	37.88 $\pm$ 0.09 & 38.45 $\pm$ 0.09 \\
\hline
	$\sigma_A^2$ (nm$^4$) & 0.35 $\pm$ 0.02 & 0.379 $\pm$ 0.015 & 0.277 $\pm$ 0.012 \\
\hline
	$\langle L\rangle$ (nm) & 7.372 $\pm$ 0.014 & 7.399 $\pm$ 0.018 & 7.290 $\pm$ 0.016 \\
\hline
	$\sigma_L^2$ (nm$^2$) & 0.0125 $\pm$ 0.0007  & 0.0143 $\pm$ 0.0006 & 0.0094 $\pm$ 0.0004\\
\end{tabular}
\end{table*}

\section{Discussions}

In this work, we introduced a barostat named stochastic cell rescaling that is driven by a first order differential equation on the volume.
The formulation makes it very similar to the popular Berendsen barostat,\cite{berendsen1984molecular} but includes a noise term to enforce
the correct volume fluctuations.
Being based on a first-order differential equation, the method can be used effectively in equilibration phases.
The tested systems range from simple fluids to macromolecular constructs
and are modeled using intra and inter-molecular interactions as well as constraints.
A version suitable to control surface tension is also presented and tested on a membrane system.
In all tested cases, the fluctuations obtained with a reference implementation of the Parrinello-Rahman algorithm were reproduced for a broad range of choices
of the relaxation time of the barostat.
Our algorithm can be implemented using either the instantaneous kinetic energy or its average value,
resulting in very similar behaviors in the tested cases. The choice between the two formulations can be guided by practical reasons,
such as the ease of their implementation in a given MD code.
The algorithm can be easily modified to make usage of the molecular virial (see, \emph{e.g.}, Refs.~\onlinecite{kalibaeva2003constant,marry2007trotter,lippert2013accurate,di2015stochastic}) provided that
molecular positions and, optionally, velocities are scaled instead of atomic ones.
Stochastic cell rescaling can in principle be combined with any thermostat to sample the isothermal-isobaric ensemble.
In particular, in the membrane simulation presented here we tested its usage in combination with the Nose-Hoover thermostat,\cite{nose1984unified,hoover1985canonical} whereas in the other simulations we used stochastic velocity rescaling.\cite{bussi2007canonical}
In general, our recommendation would be Langevin dynamics,\cite{schneider1978brownian} if a local thermostat is desired to independently thermalize all degrees of freedom,
or stochastic velocity rescaling, if a global thermostat is preferred so as to avoid slowing down particle diffusion.\cite{bussi2008stochastic}

Similarly to the Berendsen barostat, our scheme has a single control parameter, $\tau_P$, that controls the equilibration rate, but has to be complemented with
an estimate of the isothermal compressibility of the system $\beta_T$.
For the simulation of solvated systems, this compressibility is largely dependent on the properties of the solvent,
so that in typical applications one can just use an estimate of its value for water. For other systems, it can be estimated by computing volume
fluctuations $\sigma_V$ in a test run and using the relationship $\beta_T=\frac{\sigma_V^2}{\langle V\rangle k_BT}$. In any case, only the ratio $\beta_T/\tau_P$ between these two parameters
enters the algorithm, implying that an incorrect estimate of $\beta_T$ by,
for instance, a factor two would result in an error in the control
of the relaxation time of the barostat of a factor two. Given the robustness of the results as a function of $\tau_P$, we consider this
as a minor drawback. We notice that this holds also for the original Berendsen barostat.

The relationship between stochastic cell rescaling and the Berendsen barostat is very similar to
the relationship between stochastic velocity rescaling and
the Berendsen thermostat.
Importantly, the probability distribution of the kinetic energy is known \emph{a priori},
allowing the equations of the stochastic velocity rescaling thermostat to be solved exactly.\cite{bussi2007canonical}
On the contrary, the probability distribution of the volume depends in a non trivial manner on the coordinates.
As a consequence, the barostat equations must be integrated approximately with a finite time step algorithm,
and results might be incorrect if the coupling parameter $\tau_P$ is chosen too small.
We here tested a number of algorithms showing that, whereas only some of them allow detailed-balance violations to be quantified,
all of them can be used in practical applications for reasonable choices of the input parameters.
We make available all the algorithms in an educational MD code and the simplest one in the GROMACS code.
Source codes and instructions can be found at \url{http://github.com/bussilab/crescale}.

Stochastic cell rescaling can be seen as a first-order version of the Langevin piston algorithm,\cite{feller1995constant}
and indeed it can be obtained by choosing a piston inertia small enough and a friction coefficient correspondingly large
(see Supplementary Material, Sec.~II).
The analogy between the Langevin piston and the Berendsen barostat was already pointed out
in Ref.~\onlinecite{feller1995constant}, although resulting in a first-order equation that
was suggested to be difficult to implement.
Since the original Langevin piston algorithm was based on the Andersen thermostat, where absolute changes
of the volume are driven rather than relative ones, the first-order equation suggested in Ref.~\onlinecite{feller1995constant}
differs from Eq.~\ref{eq:de},
and its deterministic part thus differs from the one used in the Berendsen barostat.
The advantage of the present formulation is that it can be straightforwardly implemented in any code supporting the Berendsen barostat, whereas
a standard Langevin piston implementation should be built on top of a second-order barostat.
In addition, stochastic cell rescaling works by construction in the high-friction limit and it is thus expected to
be always stable in equilibration phases,
still retaining the correct fluctuations in production runs.

A number of papers have shown artifacts related to the use of the Berendsen thermostat, see, \emph{e.g.},
Refs.~\onlinecite{harvey1998flying,rosta2009thermostat,wong2010static,shirts2013simple,merz2018testing},
including broken reversibility and incorrect fluctuations.
The artifacts on the reversibility are relatively small.\cite{wong2010static,wong2016good}
The largest issue is the underestimation of energy fluctuations,
that has an effect on replica-exchange simulations where the incorrectly distributed energies are used
to compute acceptance probabilities.\cite{rosta2009thermostat}
The Berendsen barostat, similarly, is known to result in incorrect volume fluctuations
(see, \emph{e.g.}, Refs.~\onlinecite{shirts2013simple,rogge2015comparison,merz2018testing}),
and thus should not be used to evaluate compressibilities.
The incorrect volume fluctuations
could also introduce significant artifacts in replica-exchange simulations where replicas are simulated at different pressure\cite{okabe2001replica}
or surface tension.\cite{mori2013surface}
In these cases, a barostat reproducing correct volume fluctuations has to be considered as mandatory.
We here tested the result of free-energy calculations required to estimate ligand affinities,
that were recently reported to be dependent on barostating details and, in particular,
to display measurable artifacts induced by the use of the Berendsen barostat.\cite{rizzi2020sampl6}
According to our results, all the tested barostats, including the Berendsen one, were leading to equivalent results in this specific application.
This suggests that the discrepancies observed in Ref.~\onlinecite{rizzi2020sampl6} might be a consequence of some other implementation detail,
and that artifacts of the Berendsen barostat on the properties of solvated molecules that are not directly correlated with volume fluctuations
might be small.
Nevertheless, we recommend stochastic cell rescaling as a better alternative since, thanks to the additional noise,
it is guaranteed to sample the correct distribution.

The semi-isotropic version of stochastic cell rescaling was tested on a membrane simulation.
These tests are more qualitative, since it is difficult to obtain statistically converged results on this system
within the simulated time scales.
Further tests will be necessary to see if pathological behaviors appear when the fraction of one phase is much larger than the fraction
of the other phase, or when the difference in compressibility between the two phases is larger.
It will be important to verify how stochastic cell rescaling compares with Parrinello-Rahman\cite{parrinello1981polymorphic}
and anisotropic Martyna-Tobias-Klein\cite{martyna1994constant} algorithms in these difficult cases.
A more flexible formulation where all the elements of the cell matrix can be adjusted
will be the subject of a later work.

In summary, stochastic cell rescaling provides a simple first-order barostat that can be equally used in equilibration and production runs
and can be adopted as a drop-in replacement of the Berendsen barostat, with minimal implementation changes that allow
the isothermal-isobaric fluctuations to be properly sampled.

\section{Supplementary Material}

Supplementary Material includes Supplementary Methods related to the center-of-mass contribution
to the pressure, the relationship with Langevin piston algorithm, more details on the integration of equations
of motion and on the calculation of the energy drift, and the full derivation of the semi-isotropic
version of the barostat. Supplementary Material also includes Supplementary Results related to
the validation of the volume distributions, autocorrelation functions of the volume,
results obtained with the formulation of the barostat where the average kinetic energy is used to compute
the internal pressure, physical validation tests on a water box, and time series for the membrane simulations.

\begin{acknowledgments}
Massimiliano Bonomi, Carlo Camilloni, Michele Ceriotti, Paolo Raiteri, and Michael Shirts
are acknowledged for reading the manuscript and providing useful suggestions.
\end{acknowledgments}

\section*{Data availability}

The data that support the findings of this study are openly available
at \url{http://doi.org/10.5281/zenodo.3921886}.
Analysis scripts and modified software
can be found at \url{https://github.com/bussilab/crescale}.

\bibliography{main}%

\begin{thebibliography}{72}%
\makeatletter
\providecommand \@ifxundefined [1]{%
 \@ifx{#1\undefined}
}%
\providecommand \@ifnum [1]{%
 \ifnum #1\expandafter \@firstoftwo
 \else \expandafter \@secondoftwo
 \fi
}%
\providecommand \@ifx [1]{%
 \ifx #1\expandafter \@firstoftwo
 \else \expandafter \@secondoftwo
 \fi
}%
\providecommand \natexlab [1]{#1}%
\providecommand \enquote  [1]{``#1''}%
\providecommand \bibnamefont  [1]{#1}%
\providecommand \bibfnamefont [1]{#1}%
\providecommand \citenamefont [1]{#1}%
\providecommand \href@noop [0]{\@secondoftwo}%
\providecommand \href [0]{\begingroup \@sanitize@url \@href}%
\providecommand \@href[1]{\@@startlink{#1}\@@href}%
\providecommand \@@href[1]{\endgroup#1\@@endlink}%
\providecommand \@sanitize@url [0]{\catcode `\\12\catcode `\$12\catcode
  `\&12\catcode `\#12\catcode `\^12\catcode `\_12\catcode `\%12\relax}%
\providecommand \@@startlink[1]{}%
\providecommand \@@endlink[0]{}%
\providecommand \url  [0]{\begingroup\@sanitize@url \@url }%
\providecommand \@url [1]{\endgroup\@href {#1}{\urlprefix }}%
\providecommand \urlprefix  [0]{URL }%
\providecommand \Eprint [0]{\href }%
\providecommand \doibase [0]{http://dx.doi.org/}%
\providecommand \selectlanguage [0]{\@gobble}%
\providecommand \bibinfo  [0]{\@secondoftwo}%
\providecommand \bibfield  [0]{\@secondoftwo}%
\providecommand \translation [1]{[#1]}%
\providecommand \BibitemOpen [0]{}%
\providecommand \bibitemStop [0]{}%
\providecommand \bibitemNoStop [0]{.\EOS\space}%
\providecommand \EOS [0]{\spacefactor3000\relax}%
\providecommand \BibitemShut  [1]{\csname bibitem#1\endcsname}%
\let\auto@bib@innerbib\@empty
\bibitem [{\citenamefont {Frenkel}\ and\ \citenamefont
  {Smit}(2001)}]{frenkel2001understanding}%
  \BibitemOpen
  \bibfield  {author} {\bibinfo {author} {\bibfnamefont {D.}~\bibnamefont
  {Frenkel}}\ and\ \bibinfo {author} {\bibfnamefont {B.}~\bibnamefont {Smit}},\
  }\href@noop {} {\emph {\bibinfo {title} {Understanding molecular simulation:
  from algorithms to applications}}}\ (\bibinfo  {publisher} {Acedemic Press,
  San Diego},\ \bibinfo {year} {2001})\BibitemShut {NoStop}%
\bibitem [{\citenamefont {Andersen}(1980)}]{andersen1980molecular}%
  \BibitemOpen
  \bibfield  {author} {\bibinfo {author} {\bibfnamefont {H.~C.}\ \bibnamefont
  {Andersen}},\ }\bibfield  {title} {\enquote {\bibinfo {title} {Molecular
  dynamics simulations at constant pressure and/or temperature},}\ }\href@noop
  {} {\bibfield  {journal} {\bibinfo  {journal} {J. Chem. Phys.}\ }\textbf
  {\bibinfo {volume} {72}},\ \bibinfo {pages} {2384--2393} (\bibinfo {year}
  {1980})}\BibitemShut {NoStop}%
\bibitem [{\citenamefont {Parrinello}\ and\ \citenamefont
  {Rahman}(1980)}]{parrinello1980crystal}%
  \BibitemOpen
  \bibfield  {author} {\bibinfo {author} {\bibfnamefont {M.}~\bibnamefont
  {Parrinello}}\ and\ \bibinfo {author} {\bibfnamefont {A.}~\bibnamefont
  {Rahman}},\ }\bibfield  {title} {\enquote {\bibinfo {title} {Crystal
  structure and pair potentials: A molecular-dynamics study},}\ }\href@noop {}
  {\bibfield  {journal} {\bibinfo  {journal} {Phys. Rev. Lett.}\ }\textbf
  {\bibinfo {volume} {45}},\ \bibinfo {pages} {1196} (\bibinfo {year}
  {1980})}\BibitemShut {NoStop}%
\bibitem [{\citenamefont {Parrinello}\ and\ \citenamefont
  {Rahman}(1981)}]{parrinello1981polymorphic}%
  \BibitemOpen
  \bibfield  {author} {\bibinfo {author} {\bibfnamefont {M.}~\bibnamefont
  {Parrinello}}\ and\ \bibinfo {author} {\bibfnamefont {A.}~\bibnamefont
  {Rahman}},\ }\bibfield  {title} {\enquote {\bibinfo {title} {Polymorphic
  transitions in single crystals: A new molecular dynamics method},}\
  }\href@noop {} {\bibfield  {journal} {\bibinfo  {journal} {J. Appl. Phys.}\
  }\textbf {\bibinfo {volume} {52}},\ \bibinfo {pages} {7182--7190} (\bibinfo
  {year} {1981})}\BibitemShut {NoStop}%
\bibitem [{\citenamefont {Nos{\'e}}\ and\ \citenamefont
  {Klein}(1983)}]{nose1983constant}%
  \BibitemOpen
  \bibfield  {author} {\bibinfo {author} {\bibfnamefont {S.}~\bibnamefont
  {Nos{\'e}}}\ and\ \bibinfo {author} {\bibfnamefont {M.}~\bibnamefont
  {Klein}},\ }\bibfield  {title} {\enquote {\bibinfo {title} {Constant pressure
  molecular dynamics for molecular systems},}\ }\href@noop {} {\bibfield
  {journal} {\bibinfo  {journal} {Mol. Phys.}\ }\textbf {\bibinfo {volume}
  {50}},\ \bibinfo {pages} {1055--1076} (\bibinfo {year} {1983})}\BibitemShut
  {NoStop}%
\bibitem [{\citenamefont {Hoover}(1985)}]{hoover1985canonical}%
  \BibitemOpen
  \bibfield  {author} {\bibinfo {author} {\bibfnamefont {W.~G.}\ \bibnamefont
  {Hoover}},\ }\bibfield  {title} {\enquote {\bibinfo {title} {Canonical
  dynamics: Equilibrium phase-space distributions},}\ }\href@noop {} {\bibfield
   {journal} {\bibinfo  {journal} {Phys. Rev. A}\ }\textbf {\bibinfo {volume}
  {31}},\ \bibinfo {pages} {1695} (\bibinfo {year} {1985})}\BibitemShut
  {NoStop}%
\bibitem [{\citenamefont {Melchionna}, \citenamefont {Ciccotti},\ and\
  \citenamefont {Lee~Holian}(1993)}]{melchionna1993hoover}%
  \BibitemOpen
  \bibfield  {author} {\bibinfo {author} {\bibfnamefont {S.}~\bibnamefont
  {Melchionna}}, \bibinfo {author} {\bibfnamefont {G.}~\bibnamefont
  {Ciccotti}}, \ and\ \bibinfo {author} {\bibfnamefont {B.}~\bibnamefont
  {Lee~Holian}},\ }\bibfield  {title} {\enquote {\bibinfo {title} {Hoover {NPT}
  dynamics for systems varying in shape and size},}\ }\href@noop {} {\bibfield
  {journal} {\bibinfo  {journal} {Mol. Phys.}\ }\textbf {\bibinfo {volume}
  {78}},\ \bibinfo {pages} {533--544} (\bibinfo {year} {1993})}\BibitemShut
  {NoStop}%
\bibitem [{\citenamefont {Martyna}, \citenamefont {Tobias},\ and\ \citenamefont
  {Klein}(1994)}]{martyna1994constant}%
  \BibitemOpen
  \bibfield  {author} {\bibinfo {author} {\bibfnamefont {G.~J.}\ \bibnamefont
  {Martyna}}, \bibinfo {author} {\bibfnamefont {D.~J.}\ \bibnamefont {Tobias}},
  \ and\ \bibinfo {author} {\bibfnamefont {M.~L.}\ \bibnamefont {Klein}},\
  }\bibfield  {title} {\enquote {\bibinfo {title} {Constant pressure molecular
  dynamics algorithms},}\ }\href@noop {} {\bibfield  {journal} {\bibinfo
  {journal} {J. Chem. Phys.}\ }\textbf {\bibinfo {volume} {101}},\ \bibinfo
  {pages} {4177--4189} (\bibinfo {year} {1994})}\BibitemShut {NoStop}%
\bibitem [{\citenamefont {Zhang}\ \emph {et~al.}(1995)\citenamefont {Zhang},
  \citenamefont {Feller}, \citenamefont {Brooks},\ and\ \citenamefont
  {Pastor}}]{zhang1995computer}%
  \BibitemOpen
  \bibfield  {author} {\bibinfo {author} {\bibfnamefont {Y.}~\bibnamefont
  {Zhang}}, \bibinfo {author} {\bibfnamefont {S.~E.}\ \bibnamefont {Feller}},
  \bibinfo {author} {\bibfnamefont {B.~R.}\ \bibnamefont {Brooks}}, \ and\
  \bibinfo {author} {\bibfnamefont {R.~W.}\ \bibnamefont {Pastor}},\ }\bibfield
   {title} {\enquote {\bibinfo {title} {Computer simulation of liquid/liquid
  interfaces. i. theory and application to octane/water},}\ }\href@noop {}
  {\bibfield  {journal} {\bibinfo  {journal} {J. Chem. Phys.}\ }\textbf
  {\bibinfo {volume} {103}},\ \bibinfo {pages} {10252--10266} (\bibinfo {year}
  {1995})}\BibitemShut {NoStop}%
\bibitem [{\citenamefont {Martyna}\ \emph {et~al.}(1996)\citenamefont
  {Martyna}, \citenamefont {Tuckerman}, \citenamefont {Tobias},\ and\
  \citenamefont {Klein}}]{martyna1996explicit}%
  \BibitemOpen
  \bibfield  {author} {\bibinfo {author} {\bibfnamefont {G.~J.}\ \bibnamefont
  {Martyna}}, \bibinfo {author} {\bibfnamefont {M.~E.}\ \bibnamefont
  {Tuckerman}}, \bibinfo {author} {\bibfnamefont {D.~J.}\ \bibnamefont
  {Tobias}}, \ and\ \bibinfo {author} {\bibfnamefont {M.~L.}\ \bibnamefont
  {Klein}},\ }\bibfield  {title} {\enquote {\bibinfo {title} {Explicit
  reversible integrators for extended systems dynamics},}\ }\href@noop {}
  {\bibfield  {journal} {\bibinfo  {journal} {Mol. Phys.}\ }\textbf {\bibinfo
  {volume} {87}},\ \bibinfo {pages} {1117--1157} (\bibinfo {year}
  {1996})}\BibitemShut {NoStop}%
\bibitem [{\citenamefont {Sturgeon}\ and\ \citenamefont
  {Laird}(2000)}]{sturgeon2000symplectic}%
  \BibitemOpen
  \bibfield  {author} {\bibinfo {author} {\bibfnamefont {J.~B.}\ \bibnamefont
  {Sturgeon}}\ and\ \bibinfo {author} {\bibfnamefont {B.~B.}\ \bibnamefont
  {Laird}},\ }\bibfield  {title} {\enquote {\bibinfo {title} {Symplectic
  algorithm for constant-pressure molecular dynamics using a
  {N}os{\'e}--{P}oincar{\'e} thermostat},}\ }\href@noop {} {\bibfield
  {journal} {\bibinfo  {journal} {J. Chem. Phys.}\ }\textbf {\bibinfo {volume}
  {112}},\ \bibinfo {pages} {3474--3482} (\bibinfo {year} {2000})}\BibitemShut
  {NoStop}%
\bibitem [{\citenamefont {Kalibaeva}, \citenamefont {Ferrario},\ and\
  \citenamefont {Ciccotti}(2003)}]{kalibaeva2003constant}%
  \BibitemOpen
  \bibfield  {author} {\bibinfo {author} {\bibfnamefont {G.}~\bibnamefont
  {Kalibaeva}}, \bibinfo {author} {\bibfnamefont {M.}~\bibnamefont {Ferrario}},
  \ and\ \bibinfo {author} {\bibfnamefont {G.}~\bibnamefont {Ciccotti}},\
  }\bibfield  {title} {\enquote {\bibinfo {title} {Constant pressure-constant
  temperature molecular dynamics: a correct constrained {NPT} ensemble using
  the molecular virial},}\ }\href@noop {} {\bibfield  {journal} {\bibinfo
  {journal} {Mol. Phys.}\ }\textbf {\bibinfo {volume} {101}},\ \bibinfo {pages}
  {765--778} (\bibinfo {year} {2003})}\BibitemShut {NoStop}%
\bibitem [{\citenamefont {Marry}\ and\ \citenamefont
  {Ciccotti}(2007)}]{marry2007trotter}%
  \BibitemOpen
  \bibfield  {author} {\bibinfo {author} {\bibfnamefont {V.}~\bibnamefont
  {Marry}}\ and\ \bibinfo {author} {\bibfnamefont {G.}~\bibnamefont
  {Ciccotti}},\ }\bibfield  {title} {\enquote {\bibinfo {title} {Trotter
  derived algorithms for molecular dynamics with constraints: Velocity {V}erlet
  revisited},}\ }\href@noop {} {\bibfield  {journal} {\bibinfo  {journal} {J.
  Comput. Phys.}\ }\textbf {\bibinfo {volume} {222}},\ \bibinfo {pages}
  {428--440} (\bibinfo {year} {2007})}\BibitemShut {NoStop}%
\bibitem [{\citenamefont {Bussi}, \citenamefont {Zykova-Timan},\ and\
  \citenamefont {Parrinello}(2009)}]{bussi2009isothermal}%
  \BibitemOpen
  \bibfield  {author} {\bibinfo {author} {\bibfnamefont {G.}~\bibnamefont
  {Bussi}}, \bibinfo {author} {\bibfnamefont {T.}~\bibnamefont {Zykova-Timan}},
  \ and\ \bibinfo {author} {\bibfnamefont {M.}~\bibnamefont {Parrinello}},\
  }\bibfield  {title} {\enquote {\bibinfo {title} {Isothermal-isobaric
  molecular dynamics using stochastic velocity rescaling},}\ }\href@noop {}
  {\bibfield  {journal} {\bibinfo  {journal} {J. Chem. Phys.}\ }\textbf
  {\bibinfo {volume} {130}},\ \bibinfo {pages} {074101} (\bibinfo {year}
  {2009})}\BibitemShut {NoStop}%
\bibitem [{\citenamefont {Yu}\ \emph {et~al.}(2010)\citenamefont {Yu},
  \citenamefont {Alejandre}, \citenamefont {L{\'o}pez-Rend{\'o}n},
  \citenamefont {Martyna},\ and\ \citenamefont {Tuckerman}}]{yu2010measure}%
  \BibitemOpen
  \bibfield  {author} {\bibinfo {author} {\bibfnamefont {T.-Q.}\ \bibnamefont
  {Yu}}, \bibinfo {author} {\bibfnamefont {J.}~\bibnamefont {Alejandre}},
  \bibinfo {author} {\bibfnamefont {R.}~\bibnamefont {L{\'o}pez-Rend{\'o}n}},
  \bibinfo {author} {\bibfnamefont {G.~J.}\ \bibnamefont {Martyna}}, \ and\
  \bibinfo {author} {\bibfnamefont {M.~E.}\ \bibnamefont {Tuckerman}},\
  }\bibfield  {title} {\enquote {\bibinfo {title} {Measure-preserving
  integrators for molecular dynamics in the isothermal--isobaric ensemble
  derived from the liouville operator},}\ }\href@noop {} {\bibfield  {journal}
  {\bibinfo  {journal} {Chem. Phys.}\ }\textbf {\bibinfo {volume} {370}},\
  \bibinfo {pages} {294--305} (\bibinfo {year} {2010})}\BibitemShut {NoStop}%
\bibitem [{\citenamefont {Raiteri}, \citenamefont {Gale},\ and\ \citenamefont
  {Bussi}(2011)}]{raiteri2011reactive}%
  \BibitemOpen
  \bibfield  {author} {\bibinfo {author} {\bibfnamefont {P.}~\bibnamefont
  {Raiteri}}, \bibinfo {author} {\bibfnamefont {J.~D.}\ \bibnamefont {Gale}}, \
  and\ \bibinfo {author} {\bibfnamefont {G.}~\bibnamefont {Bussi}},\ }\bibfield
   {title} {\enquote {\bibinfo {title} {Reactive force field simulation of
  proton diffusion in {BaZrO3} using an empirical valence bond approach},}\
  }\href@noop {} {\bibfield  {journal} {\bibinfo  {journal} {J. Phys. Condens.
  Matter}\ }\textbf {\bibinfo {volume} {23}},\ \bibinfo {pages} {334213}
  (\bibinfo {year} {2011})}\BibitemShut {NoStop}%
\bibitem [{\citenamefont {Lippert}\ \emph {et~al.}(2013)\citenamefont
  {Lippert}, \citenamefont {Predescu}, \citenamefont {Ierardi}, \citenamefont
  {Mackenzie}, \citenamefont {Eastwood}, \citenamefont {Dror},\ and\
  \citenamefont {Shaw}}]{lippert2013accurate}%
  \BibitemOpen
  \bibfield  {author} {\bibinfo {author} {\bibfnamefont {R.~A.}\ \bibnamefont
  {Lippert}}, \bibinfo {author} {\bibfnamefont {C.}~\bibnamefont {Predescu}},
  \bibinfo {author} {\bibfnamefont {D.~J.}\ \bibnamefont {Ierardi}}, \bibinfo
  {author} {\bibfnamefont {K.~M.}\ \bibnamefont {Mackenzie}}, \bibinfo {author}
  {\bibfnamefont {M.~P.}\ \bibnamefont {Eastwood}}, \bibinfo {author}
  {\bibfnamefont {R.~O.}\ \bibnamefont {Dror}}, \ and\ \bibinfo {author}
  {\bibfnamefont {D.~E.}\ \bibnamefont {Shaw}},\ }\bibfield  {title} {\enquote
  {\bibinfo {title} {Accurate and efficient integration for molecular dynamics
  simulations at constant temperature and pressure},}\ }\href@noop {}
  {\bibfield  {journal} {\bibinfo  {journal} {J. Chem. Phys.}\ }\textbf
  {\bibinfo {volume} {139}},\ \bibinfo {pages} {164106} (\bibinfo {year}
  {2013})}\BibitemShut {NoStop}%
\bibitem [{\citenamefont {Feller}\ \emph {et~al.}(1995)\citenamefont {Feller},
  \citenamefont {Zhang}, \citenamefont {Pastor},\ and\ \citenamefont
  {Brooks}}]{feller1995constant}%
  \BibitemOpen
  \bibfield  {author} {\bibinfo {author} {\bibfnamefont {S.~E.}\ \bibnamefont
  {Feller}}, \bibinfo {author} {\bibfnamefont {Y.}~\bibnamefont {Zhang}},
  \bibinfo {author} {\bibfnamefont {R.~W.}\ \bibnamefont {Pastor}}, \ and\
  \bibinfo {author} {\bibfnamefont {B.~R.}\ \bibnamefont {Brooks}},\ }\bibfield
   {title} {\enquote {\bibinfo {title} {Constant pressure molecular dynamics
  simulation: the {L}angevin piston method},}\ }\href@noop {} {\bibfield
  {journal} {\bibinfo  {journal} {J. Chem. Phys.}\ }\textbf {\bibinfo {volume}
  {103}},\ \bibinfo {pages} {4613--4621} (\bibinfo {year} {1995})}\BibitemShut
  {NoStop}%
\bibitem [{\citenamefont {Kolb}\ and\ \citenamefont
  {D{\"u}nweg}(1999)}]{kolb1999optimized}%
  \BibitemOpen
  \bibfield  {author} {\bibinfo {author} {\bibfnamefont {A.}~\bibnamefont
  {Kolb}}\ and\ \bibinfo {author} {\bibfnamefont {B.}~\bibnamefont
  {D{\"u}nweg}},\ }\bibfield  {title} {\enquote {\bibinfo {title} {Optimized
  constant pressure stochastic dynamics},}\ }\href@noop {} {\bibfield
  {journal} {\bibinfo  {journal} {J. Chem. Phys.}\ }\textbf {\bibinfo {volume}
  {111}},\ \bibinfo {pages} {4453--4459} (\bibinfo {year} {1999})}\BibitemShut
  {NoStop}%
\bibitem [{\citenamefont {Quigley}\ and\ \citenamefont
  {Probert}(2004)}]{quigley2004langevin}%
  \BibitemOpen
  \bibfield  {author} {\bibinfo {author} {\bibfnamefont {D.}~\bibnamefont
  {Quigley}}\ and\ \bibinfo {author} {\bibfnamefont {M.}~\bibnamefont
  {Probert}},\ }\bibfield  {title} {\enquote {\bibinfo {title} {Langevin
  dynamics in constant pressure extended systems},}\ }\href@noop {} {\bibfield
  {journal} {\bibinfo  {journal} {J. Chem. Phys.}\ }\textbf {\bibinfo {volume}
  {120}},\ \bibinfo {pages} {11432--11441} (\bibinfo {year}
  {2004})}\BibitemShut {NoStop}%
\bibitem [{\citenamefont {Gr{\o}nbech-Jensen}\ and\ \citenamefont
  {Farago}(2014)}]{gronbech2014constant}%
  \BibitemOpen
  \bibfield  {author} {\bibinfo {author} {\bibfnamefont {N.}~\bibnamefont
  {Gr{\o}nbech-Jensen}}\ and\ \bibinfo {author} {\bibfnamefont
  {O.}~\bibnamefont {Farago}},\ }\bibfield  {title} {\enquote {\bibinfo {title}
  {Constant pressure and temperature discrete-time {L}angevin molecular
  dynamics},}\ }\href@noop {} {\bibfield  {journal} {\bibinfo  {journal} {J.
  Chem. Phys.}\ }\textbf {\bibinfo {volume} {141}},\ \bibinfo {pages} {194108}
  (\bibinfo {year} {2014})}\BibitemShut {NoStop}%
\bibitem [{\citenamefont {Di~Pierro}, \citenamefont {Elber},\ and\
  \citenamefont {Leimkuhler}(2015)}]{di2015stochastic}%
  \BibitemOpen
  \bibfield  {author} {\bibinfo {author} {\bibfnamefont {M.}~\bibnamefont
  {Di~Pierro}}, \bibinfo {author} {\bibfnamefont {R.}~\bibnamefont {Elber}}, \
  and\ \bibinfo {author} {\bibfnamefont {B.}~\bibnamefont {Leimkuhler}},\
  }\bibfield  {title} {\enquote {\bibinfo {title} {A stochastic algorithm for
  the isobaric--isothermal ensemble with {E}wald summations for all long range
  forces},}\ }\href@noop {} {\bibfield  {journal} {\bibinfo  {journal} {J.
  Chem. Theory Comput.}\ }\textbf {\bibinfo {volume} {11}},\ \bibinfo {pages}
  {5624--5637} (\bibinfo {year} {2015})}\BibitemShut {NoStop}%
\bibitem [{\citenamefont {Gao}, \citenamefont {Fang},\ and\ \citenamefont
  {Wang}(2016)}]{gao2016sampling}%
  \BibitemOpen
  \bibfield  {author} {\bibinfo {author} {\bibfnamefont {X.}~\bibnamefont
  {Gao}}, \bibinfo {author} {\bibfnamefont {J.}~\bibnamefont {Fang}}, \ and\
  \bibinfo {author} {\bibfnamefont {H.}~\bibnamefont {Wang}},\ }\bibfield
  {title} {\enquote {\bibinfo {title} {Sampling the isothermal-isobaric
  ensemble by {L}angevin dynamics},}\ }\href@noop {} {\bibfield  {journal}
  {\bibinfo  {journal} {J. Chem. Phys.}\ }\textbf {\bibinfo {volume} {144}},\
  \bibinfo {pages} {124113} (\bibinfo {year} {2016})}\BibitemShut {NoStop}%
\bibitem [{\citenamefont {Cajahuaringa}\ and\ \citenamefont
  {Antonelli}(2018)}]{cajahuaringa2018stochastic}%
  \BibitemOpen
  \bibfield  {author} {\bibinfo {author} {\bibfnamefont {S.}~\bibnamefont
  {Cajahuaringa}}\ and\ \bibinfo {author} {\bibfnamefont {A.}~\bibnamefont
  {Antonelli}},\ }\bibfield  {title} {\enquote {\bibinfo {title} {Stochastic
  sampling of the isothermal-isobaric ensemble: Phase diagram of crystalline
  solids from molecular dynamics simulation},}\ }\href@noop {} {\bibfield
  {journal} {\bibinfo  {journal} {J. Chem. Phys.}\ }\textbf {\bibinfo {volume}
  {149}},\ \bibinfo {pages} {064114} (\bibinfo {year} {2018})}\BibitemShut
  {NoStop}%
\bibitem [{\citenamefont {Chow}\ and\ \citenamefont
  {Ferguson}(1995)}]{chow1995isothermal}%
  \BibitemOpen
  \bibfield  {author} {\bibinfo {author} {\bibfnamefont {K.-H.}\ \bibnamefont
  {Chow}}\ and\ \bibinfo {author} {\bibfnamefont {D.~M.}\ \bibnamefont
  {Ferguson}},\ }\bibfield  {title} {\enquote {\bibinfo {title}
  {Isothermal-isobaric molecular dynamics simulations with {M}onte {C}arlo
  volume sampling},}\ }\href@noop {} {\bibfield  {journal} {\bibinfo  {journal}
  {Comput. Phys. Commun.}\ }\textbf {\bibinfo {volume} {91}},\ \bibinfo {pages}
  {283--289} (\bibinfo {year} {1995})}\BibitemShut {NoStop}%
\bibitem [{\citenamefont {{\AA}qvist}\ \emph {et~al.}(2004)\citenamefont
  {{\AA}qvist}, \citenamefont {Wennerstr{\"o}m}, \citenamefont {Nervall},
  \citenamefont {Bjelic},\ and\ \citenamefont
  {Brandsdal}}]{aaqvist2004molecular}%
  \BibitemOpen
  \bibfield  {author} {\bibinfo {author} {\bibfnamefont {J.}~\bibnamefont
  {{\AA}qvist}}, \bibinfo {author} {\bibfnamefont {P.}~\bibnamefont
  {Wennerstr{\"o}m}}, \bibinfo {author} {\bibfnamefont {M.}~\bibnamefont
  {Nervall}}, \bibinfo {author} {\bibfnamefont {S.}~\bibnamefont {Bjelic}}, \
  and\ \bibinfo {author} {\bibfnamefont {B.~O.}\ \bibnamefont {Brandsdal}},\
  }\bibfield  {title} {\enquote {\bibinfo {title} {Molecular dynamics
  simulations of water and biomolecules with a {M}onte {C}arlo constant
  pressure algorithm},}\ }\href@noop {} {\bibfield  {journal} {\bibinfo
  {journal} {Chem. Phys. Lett.}\ }\textbf {\bibinfo {volume} {384}},\ \bibinfo
  {pages} {288--294} (\bibinfo {year} {2004})}\BibitemShut {NoStop}%
\bibitem [{\citenamefont {Harger}\ and\ \citenamefont
  {Ren}(2019)}]{harger2019virial}%
  \BibitemOpen
  \bibfield  {author} {\bibinfo {author} {\bibfnamefont {M.}~\bibnamefont
  {Harger}}\ and\ \bibinfo {author} {\bibfnamefont {P.}~\bibnamefont {Ren}},\
  }\bibfield  {title} {\enquote {\bibinfo {title} {Virial-based {B}erendsen
  barostat on {GPU}s using {AMOEBA} in {Tinker-OpenMM}},}\ }\href@noop {}
  {\bibfield  {journal} {\bibinfo  {journal} {Results Chem.}\ }\textbf
  {\bibinfo {volume} {1}},\ \bibinfo {pages} {100004} (\bibinfo {year}
  {2019})}\BibitemShut {NoStop}%
\bibitem [{\citenamefont {Berendsen}\ \emph {et~al.}(1984)\citenamefont
  {Berendsen}, \citenamefont {Postma}, \citenamefont {van Gunsteren},
  \citenamefont {DiNola},\ and\ \citenamefont {Haak}}]{berendsen1984molecular}%
  \BibitemOpen
  \bibfield  {author} {\bibinfo {author} {\bibfnamefont {H.~J.}\ \bibnamefont
  {Berendsen}}, \bibinfo {author} {\bibfnamefont {J.~v.}\ \bibnamefont
  {Postma}}, \bibinfo {author} {\bibfnamefont {W.~F.}\ \bibnamefont {van
  Gunsteren}}, \bibinfo {author} {\bibfnamefont {A.}~\bibnamefont {DiNola}}, \
  and\ \bibinfo {author} {\bibfnamefont {J.~R.}\ \bibnamefont {Haak}},\
  }\bibfield  {title} {\enquote {\bibinfo {title} {Molecular dynamics with
  coupling to an external bath},}\ }\href@noop {} {\bibfield  {journal}
  {\bibinfo  {journal} {J. Chem. Phys.}\ }\textbf {\bibinfo {volume} {81}},\
  \bibinfo {pages} {3684--3690} (\bibinfo {year} {1984})}\BibitemShut {NoStop}%
\bibitem [{\citenamefont {Braun}\ \emph {et~al.}(2019)\citenamefont {Braun},
  \citenamefont {Gilmer}, \citenamefont {Mayes}, \citenamefont {Mobley},
  \citenamefont {Monroe}, \citenamefont {Prasad},\ and\ \citenamefont
  {Zuckerman}}]{braun2019best}%
  \BibitemOpen
  \bibfield  {author} {\bibinfo {author} {\bibfnamefont {E.}~\bibnamefont
  {Braun}}, \bibinfo {author} {\bibfnamefont {J.}~\bibnamefont {Gilmer}},
  \bibinfo {author} {\bibfnamefont {H.~B.}\ \bibnamefont {Mayes}}, \bibinfo
  {author} {\bibfnamefont {D.~L.}\ \bibnamefont {Mobley}}, \bibinfo {author}
  {\bibfnamefont {J.~I.}\ \bibnamefont {Monroe}}, \bibinfo {author}
  {\bibfnamefont {S.}~\bibnamefont {Prasad}}, \ and\ \bibinfo {author}
  {\bibfnamefont {D.~M.}\ \bibnamefont {Zuckerman}},\ }\bibfield  {title}
  {\enquote {\bibinfo {title} {Best practices for foundations in molecular
  simulations [article v1. 0]},}\ }\href@noop {} {\bibfield  {journal}
  {\bibinfo  {journal} {Living J. Comp. Mol. Sci.}\ }\textbf {\bibinfo {volume}
  {1}},\ \bibinfo {pages} {5957} (\bibinfo {year} {2019})}\BibitemShut
  {NoStop}%
\bibitem [{\citenamefont {Bussi}, \citenamefont {Donadio},\ and\ \citenamefont
  {Parrinello}(2007)}]{bussi2007canonical}%
  \BibitemOpen
  \bibfield  {author} {\bibinfo {author} {\bibfnamefont {G.}~\bibnamefont
  {Bussi}}, \bibinfo {author} {\bibfnamefont {D.}~\bibnamefont {Donadio}}, \
  and\ \bibinfo {author} {\bibfnamefont {M.}~\bibnamefont {Parrinello}},\
  }\bibfield  {title} {\enquote {\bibinfo {title} {Canonical sampling through
  velocity rescaling},}\ }\href@noop {} {\bibfield  {journal} {\bibinfo
  {journal} {J. Chem. Phys.}\ }\textbf {\bibinfo {volume} {126}},\ \bibinfo
  {pages} {014101} (\bibinfo {year} {2007})}\BibitemShut {NoStop}%
\bibitem [{\citenamefont {Abraham}\ \emph {et~al.}(2015)\citenamefont
  {Abraham}, \citenamefont {Murtola}, \citenamefont {Schulz}, \citenamefont
  {P{\'a}ll}, \citenamefont {Smith}, \citenamefont {Hess},\ and\ \citenamefont
  {Lindahl}}]{abraham2015gromacs}%
  \BibitemOpen
  \bibfield  {author} {\bibinfo {author} {\bibfnamefont {M.~J.}\ \bibnamefont
  {Abraham}}, \bibinfo {author} {\bibfnamefont {T.}~\bibnamefont {Murtola}},
  \bibinfo {author} {\bibfnamefont {R.}~\bibnamefont {Schulz}}, \bibinfo
  {author} {\bibfnamefont {S.}~\bibnamefont {P{\'a}ll}}, \bibinfo {author}
  {\bibfnamefont {J.~C.}\ \bibnamefont {Smith}}, \bibinfo {author}
  {\bibfnamefont {B.}~\bibnamefont {Hess}}, \ and\ \bibinfo {author}
  {\bibfnamefont {E.}~\bibnamefont {Lindahl}},\ }\bibfield  {title} {\enquote
  {\bibinfo {title} {{GROMACS}: High performance molecular simulations through
  multi-level parallelism from laptops to supercomputers},}\ }\href@noop {}
  {\bibfield  {journal} {\bibinfo  {journal} {SoftwareX}\ }\textbf {\bibinfo
  {volume} {1--2}},\ \bibinfo {pages} {19--25} (\bibinfo {year}
  {2015})}\BibitemShut {NoStop}%
\bibitem [{\citenamefont {Rizzi}\ \emph {et~al.}(2020)\citenamefont {Rizzi},
  \citenamefont {Jensen}, \citenamefont {Slochower}, \citenamefont {Aldeghi},
  \citenamefont {Gapsys}, \citenamefont {Ntekoumes}, \citenamefont {Bosisio},
  \citenamefont {Papadourakis}, \citenamefont {Henriksen}, \citenamefont
  {De~Groot}, \citenamefont {Cournia}, \citenamefont {Dickson}, \citenamefont
  {Michel}, \citenamefont {Gilson}, \citenamefont {Shirts}, \citenamefont
  {Mobley},\ and\ \citenamefont {Chodera}}]{rizzi2020sampl6}%
  \BibitemOpen
  \bibfield  {author} {\bibinfo {author} {\bibfnamefont {A.}~\bibnamefont
  {Rizzi}}, \bibinfo {author} {\bibfnamefont {T.}~\bibnamefont {Jensen}},
  \bibinfo {author} {\bibfnamefont {D.~R.}\ \bibnamefont {Slochower}}, \bibinfo
  {author} {\bibfnamefont {M.}~\bibnamefont {Aldeghi}}, \bibinfo {author}
  {\bibfnamefont {V.}~\bibnamefont {Gapsys}}, \bibinfo {author} {\bibfnamefont
  {D.}~\bibnamefont {Ntekoumes}}, \bibinfo {author} {\bibfnamefont
  {S.}~\bibnamefont {Bosisio}}, \bibinfo {author} {\bibfnamefont
  {M.}~\bibnamefont {Papadourakis}}, \bibinfo {author} {\bibfnamefont {N.~M.}\
  \bibnamefont {Henriksen}}, \bibinfo {author} {\bibfnamefont {B.~L.}\
  \bibnamefont {De~Groot}}, \bibinfo {author} {\bibfnamefont {Z.}~\bibnamefont
  {Cournia}}, \bibinfo {author} {\bibfnamefont {A.}~\bibnamefont {Dickson}},
  \bibinfo {author} {\bibfnamefont {J.}~\bibnamefont {Michel}}, \bibinfo
  {author} {\bibfnamefont {M.~K.}\ \bibnamefont {Gilson}}, \bibinfo {author}
  {\bibfnamefont {M.~R.}\ \bibnamefont {Shirts}}, \bibinfo {author}
  {\bibfnamefont {D.~L.}\ \bibnamefont {Mobley}}, \ and\ \bibinfo {author}
  {\bibfnamefont {J.~D.}\ \bibnamefont {Chodera}},\ }\bibfield  {title}
  {\enquote {\bibinfo {title} {The {SAMPL}6 sampling challenge: Assessing the
  reliability and efficiency of binding free energy calculations},}\
  }\href@noop {} {\bibfield  {journal} {\bibinfo  {journal} {J. Comput. Aided
  Mol. Des.}\ }\textbf {\bibinfo {volume} {32}},\ \bibinfo {pages} {610--633}
  (\bibinfo {year} {2020})}\BibitemShut {NoStop}%
\bibitem [{\citenamefont {Gardiner}(2009)}]{gardiner2009handbook}%
  \BibitemOpen
  \bibfield  {author} {\bibinfo {author} {\bibfnamefont {C.~W.}\ \bibnamefont
  {Gardiner}},\ }\href@noop {} {\emph {\bibinfo {title} {Handbook of stochastic
  methods}}}\ (\bibinfo  {publisher} {Springer Berlin},\ \bibinfo {year}
  {2009})\BibitemShut {NoStop}%
\bibitem [{\citenamefont {Matta}\ \emph {et~al.}(2011)\citenamefont {Matta},
  \citenamefont {Massa}, \citenamefont {Gubskaya},\ and\ \citenamefont
  {Knoll}}]{matta2011can}%
  \BibitemOpen
  \bibfield  {author} {\bibinfo {author} {\bibfnamefont {C.~F.}\ \bibnamefont
  {Matta}}, \bibinfo {author} {\bibfnamefont {L.}~\bibnamefont {Massa}},
  \bibinfo {author} {\bibfnamefont {A.~V.}\ \bibnamefont {Gubskaya}}, \ and\
  \bibinfo {author} {\bibfnamefont {E.}~\bibnamefont {Knoll}},\ }\bibfield
  {title} {\enquote {\bibinfo {title} {Can one take the logarithm or the sine
  of a dimensioned quantity or a unit? dimensional analysis involving
  transcendental functions},}\ }\href@noop {} {\bibfield  {journal} {\bibinfo
  {journal} {J. Chem. Educ.}\ }\textbf {\bibinfo {volume} {88}},\ \bibinfo
  {pages} {67--70} (\bibinfo {year} {2011})}\BibitemShut {NoStop}%
\bibitem [{\citenamefont {Hottovy}\ \emph {et~al.}(2015)\citenamefont
  {Hottovy}, \citenamefont {McDaniel}, \citenamefont {Volpe},\ and\
  \citenamefont {Wehr}}]{hottovy2015smoluchowski}%
  \BibitemOpen
  \bibfield  {author} {\bibinfo {author} {\bibfnamefont {S.}~\bibnamefont
  {Hottovy}}, \bibinfo {author} {\bibfnamefont {A.}~\bibnamefont {McDaniel}},
  \bibinfo {author} {\bibfnamefont {G.}~\bibnamefont {Volpe}}, \ and\ \bibinfo
  {author} {\bibfnamefont {J.}~\bibnamefont {Wehr}},\ }\bibfield  {title}
  {\enquote {\bibinfo {title} {The {S}moluchowski-{K}ramers limit of stochastic
  differential equations with arbitrary state-dependent friction},}\
  }\href@noop {} {\bibfield  {journal} {\bibinfo  {journal} {Commun. Math.
  Phys}\ }\textbf {\bibinfo {volume} {336}},\ \bibinfo {pages} {1259--1283}
  (\bibinfo {year} {2015})}\BibitemShut {NoStop}%
\bibitem [{\citenamefont {Ryckaert}, \citenamefont {Ciccotti},\ and\
  \citenamefont {Berendsen}(1977)}]{ryckaert1977numerical}%
  \BibitemOpen
  \bibfield  {author} {\bibinfo {author} {\bibfnamefont {J.-P.}\ \bibnamefont
  {Ryckaert}}, \bibinfo {author} {\bibfnamefont {G.}~\bibnamefont {Ciccotti}},
  \ and\ \bibinfo {author} {\bibfnamefont {H.~J.}\ \bibnamefont {Berendsen}},\
  }\bibfield  {title} {\enquote {\bibinfo {title} {Numerical integration of the
  cartesian equations of motion of a system with constraints: molecular
  dynamics of n-alkanes},}\ }\href@noop {} {\bibfield  {journal} {\bibinfo
  {journal} {J. Comput. Phys.}\ }\textbf {\bibinfo {volume} {23}},\ \bibinfo
  {pages} {327--341} (\bibinfo {year} {1977})}\BibitemShut {NoStop}%
\bibitem [{\citenamefont {Hess}\ \emph {et~al.}(1997)\citenamefont {Hess},
  \citenamefont {Bekker}, \citenamefont {Berendsen},\ and\ \citenamefont
  {Fraaije}}]{hess1997lincs}%
  \BibitemOpen
  \bibfield  {author} {\bibinfo {author} {\bibfnamefont {B.}~\bibnamefont
  {Hess}}, \bibinfo {author} {\bibfnamefont {H.}~\bibnamefont {Bekker}},
  \bibinfo {author} {\bibfnamefont {H.~J.}\ \bibnamefont {Berendsen}}, \ and\
  \bibinfo {author} {\bibfnamefont {J.~G.}\ \bibnamefont {Fraaije}},\
  }\bibfield  {title} {\enquote {\bibinfo {title} {{LINCS}: a linear constraint
  solver for molecular simulations},}\ }\href@noop {} {\bibfield  {journal}
  {\bibinfo  {journal} {J. Comput. Chem.}\ }\textbf {\bibinfo {volume} {18}},\
  \bibinfo {pages} {1463--1472} (\bibinfo {year} {1997})}\BibitemShut {NoStop}%
\bibitem [{\citenamefont {Miyamoto}\ and\ \citenamefont
  {Kollman}(1992)}]{miyamoto1992settle}%
  \BibitemOpen
  \bibfield  {author} {\bibinfo {author} {\bibfnamefont {S.}~\bibnamefont
  {Miyamoto}}\ and\ \bibinfo {author} {\bibfnamefont {P.~A.}\ \bibnamefont
  {Kollman}},\ }\bibfield  {title} {\enquote {\bibinfo {title} {Settle: An
  analytical version of the {SHAKE} and {RATTLE} algorithm for rigid water
  models},}\ }\href@noop {} {\bibfield  {journal} {\bibinfo  {journal} {J.
  Comput. Chem.}\ }\textbf {\bibinfo {volume} {13}},\ \bibinfo {pages}
  {952--962} (\bibinfo {year} {1992})}\BibitemShut {NoStop}%
\bibitem [{\citenamefont {Bussi}\ and\ \citenamefont
  {Parrinello}(2007)}]{bussi2007accurate}%
  \BibitemOpen
  \bibfield  {author} {\bibinfo {author} {\bibfnamefont {G.}~\bibnamefont
  {Bussi}}\ and\ \bibinfo {author} {\bibfnamefont {M.}~\bibnamefont
  {Parrinello}},\ }\bibfield  {title} {\enquote {\bibinfo {title} {Accurate
  sampling using {L}angevin dynamics},}\ }\href@noop {} {\bibfield  {journal}
  {\bibinfo  {journal} {Phys. Rev. E}\ }\textbf {\bibinfo {volume} {75}},\
  \bibinfo {pages} {056707} (\bibinfo {year} {2007})}\BibitemShut {NoStop}%
\bibitem [{\citenamefont {Scemama}\ \emph {et~al.}(2006)\citenamefont
  {Scemama}, \citenamefont {Leli{\`e}vre}, \citenamefont {Stoltz},
  \citenamefont {Canc{\`e}s},\ and\ \citenamefont
  {Caffarel}}]{scemama2006efficient}%
  \BibitemOpen
  \bibfield  {author} {\bibinfo {author} {\bibfnamefont {A.}~\bibnamefont
  {Scemama}}, \bibinfo {author} {\bibfnamefont {T.}~\bibnamefont
  {Leli{\`e}vre}}, \bibinfo {author} {\bibfnamefont {G.}~\bibnamefont
  {Stoltz}}, \bibinfo {author} {\bibfnamefont {E.}~\bibnamefont {Canc{\`e}s}},
  \ and\ \bibinfo {author} {\bibfnamefont {M.}~\bibnamefont {Caffarel}},\
  }\bibfield  {title} {\enquote {\bibinfo {title} {An efficient sampling
  algorithm for variational {M}onte {C}arlo},}\ }\href@noop {} {\bibfield
  {journal} {\bibinfo  {journal} {J. Chem. Phys.}\ }\textbf {\bibinfo {volume}
  {125}},\ \bibinfo {pages} {114105} (\bibinfo {year} {2006})}\BibitemShut
  {NoStop}%
\bibitem [{\citenamefont {Sivak}, \citenamefont {Chodera},\ and\ \citenamefont
  {Crooks}(2013)}]{sivak2013using}%
  \BibitemOpen
  \bibfield  {author} {\bibinfo {author} {\bibfnamefont {D.~A.}\ \bibnamefont
  {Sivak}}, \bibinfo {author} {\bibfnamefont {J.~D.}\ \bibnamefont {Chodera}},
  \ and\ \bibinfo {author} {\bibfnamefont {G.~E.}\ \bibnamefont {Crooks}},\
  }\bibfield  {title} {\enquote {\bibinfo {title} {Using nonequilibrium
  fluctuation theorems to understand and correct errors in equilibrium and
  nonequilibrium simulations of discrete {L}angevin dynamics},}\ }\href@noop {}
  {\bibfield  {journal} {\bibinfo  {journal} {Phys. Rev. X}\ }\textbf {\bibinfo
  {volume} {3}},\ \bibinfo {pages} {011007} (\bibinfo {year}
  {2013})}\BibitemShut {NoStop}%
\bibitem [{\citenamefont {Rossky}, \citenamefont {Doll},\ and\ \citenamefont
  {Friedman}(1978)}]{rossky1978brownian}%
  \BibitemOpen
  \bibfield  {author} {\bibinfo {author} {\bibfnamefont {P.~J.}\ \bibnamefont
  {Rossky}}, \bibinfo {author} {\bibfnamefont {J.}~\bibnamefont {Doll}}, \ and\
  \bibinfo {author} {\bibfnamefont {H.}~\bibnamefont {Friedman}},\ }\bibfield
  {title} {\enquote {\bibinfo {title} {Brownian dynamics as smart {M}onte
  {C}arlo simulation},}\ }\href@noop {} {\bibfield  {journal} {\bibinfo
  {journal} {J. Chem. Phys.}\ }\textbf {\bibinfo {volume} {69}},\ \bibinfo
  {pages} {4628--4633} (\bibinfo {year} {1978})}\BibitemShut {NoStop}%
\bibitem [{\citenamefont {Manousiouthakis}\ and\ \citenamefont
  {Deem}(1999)}]{manousiouthakis1999strict}%
  \BibitemOpen
  \bibfield  {author} {\bibinfo {author} {\bibfnamefont {V.~I.}\ \bibnamefont
  {Manousiouthakis}}\ and\ \bibinfo {author} {\bibfnamefont {M.~W.}\
  \bibnamefont {Deem}},\ }\bibfield  {title} {\enquote {\bibinfo {title}
  {Strict detailed balance is unnecessary in {M}onte {C}arlo simulation},}\
  }\href@noop {} {\bibfield  {journal} {\bibinfo  {journal} {J. Chem. Phys.}\
  }\textbf {\bibinfo {volume} {110}},\ \bibinfo {pages} {2753--2756} (\bibinfo
  {year} {1999})}\BibitemShut {NoStop}%
\bibitem [{\citenamefont {Fass}\ \emph {et~al.}(2018)\citenamefont {Fass},
  \citenamefont {Sivak}, \citenamefont {Crooks}, \citenamefont {Beauchamp},
  \citenamefont {Leimkuhler},\ and\ \citenamefont
  {Chodera}}]{fass2018quantifying}%
  \BibitemOpen
  \bibfield  {author} {\bibinfo {author} {\bibfnamefont {J.}~\bibnamefont
  {Fass}}, \bibinfo {author} {\bibfnamefont {D.~A.}\ \bibnamefont {Sivak}},
  \bibinfo {author} {\bibfnamefont {G.~E.}\ \bibnamefont {Crooks}}, \bibinfo
  {author} {\bibfnamefont {K.~A.}\ \bibnamefont {Beauchamp}}, \bibinfo {author}
  {\bibfnamefont {B.}~\bibnamefont {Leimkuhler}}, \ and\ \bibinfo {author}
  {\bibfnamefont {J.~D.}\ \bibnamefont {Chodera}},\ }\bibfield  {title}
  {\enquote {\bibinfo {title} {Quantifying configuration-sampling error in
  {L}angevin simulations of complex molecular systems},}\ }\href@noop {}
  {\bibfield  {journal} {\bibinfo  {journal} {Entropy}\ }\textbf {\bibinfo
  {volume} {20}},\ \bibinfo {pages} {318} (\bibinfo {year} {2018})}\BibitemShut
  {NoStop}%
\bibitem [{\citenamefont {Tuckerman}, \citenamefont {Martyna},\ and\
  \citenamefont {Berne}(1990)}]{tuckerman1990molecular}%
  \BibitemOpen
  \bibfield  {author} {\bibinfo {author} {\bibfnamefont {M.~E.}\ \bibnamefont
  {Tuckerman}}, \bibinfo {author} {\bibfnamefont {G.~J.}\ \bibnamefont
  {Martyna}}, \ and\ \bibinfo {author} {\bibfnamefont {B.~J.}\ \bibnamefont
  {Berne}},\ }\bibfield  {title} {\enquote {\bibinfo {title} {Molecular
  dynamics algorithm for condensed systems with multiple time scales},}\
  }\href@noop {} {\bibfield  {journal} {\bibinfo  {journal} {J. Chem. Phys.}\
  }\textbf {\bibinfo {volume} {93}},\ \bibinfo {pages} {1287--1291} (\bibinfo
  {year} {1990})}\BibitemShut {NoStop}%
\bibitem [{\citenamefont {Jorgensen}\ \emph {et~al.}(1983)\citenamefont
  {Jorgensen}, \citenamefont {Chandrasekhar}, \citenamefont {Madura},
  \citenamefont {Impey},\ and\ \citenamefont
  {Klein}}]{jorgensen1983comparison}%
  \BibitemOpen
  \bibfield  {author} {\bibinfo {author} {\bibfnamefont {W.~L.}\ \bibnamefont
  {Jorgensen}}, \bibinfo {author} {\bibfnamefont {J.}~\bibnamefont
  {Chandrasekhar}}, \bibinfo {author} {\bibfnamefont {J.~D.}\ \bibnamefont
  {Madura}}, \bibinfo {author} {\bibfnamefont {R.~W.}\ \bibnamefont {Impey}}, \
  and\ \bibinfo {author} {\bibfnamefont {M.~L.}\ \bibnamefont {Klein}},\
  }\bibfield  {title} {\enquote {\bibinfo {title} {Comparison of simple
  potential functions for simulating liquid water},}\ }\href@noop {} {\bibfield
   {journal} {\bibinfo  {journal} {J. Chem. Phys.}\ }\textbf {\bibinfo {volume}
  {79}},\ \bibinfo {pages} {926--935} (\bibinfo {year} {1983})}\BibitemShut
  {NoStop}%
\bibitem [{\citenamefont {Shirts}(2013)}]{shirts2013simple}%
  \BibitemOpen
  \bibfield  {author} {\bibinfo {author} {\bibfnamefont {M.~R.}\ \bibnamefont
  {Shirts}},\ }\bibfield  {title} {\enquote {\bibinfo {title} {Simple
  quantitative tests to validate sampling from thermodynamic ensembles},}\
  }\href@noop {} {\bibfield  {journal} {\bibinfo  {journal} {J. Chem. Theory
  Comput.}\ }\textbf {\bibinfo {volume} {9}},\ \bibinfo {pages} {909--926}
  (\bibinfo {year} {2013})}\BibitemShut {NoStop}%
\bibitem [{\citenamefont {Merz}\ and\ \citenamefont
  {Shirts}(2018)}]{merz2018testing}%
  \BibitemOpen
  \bibfield  {author} {\bibinfo {author} {\bibfnamefont {P.~T.}\ \bibnamefont
  {Merz}}\ and\ \bibinfo {author} {\bibfnamefont {M.~R.}\ \bibnamefont
  {Shirts}},\ }\bibfield  {title} {\enquote {\bibinfo {title} {Testing for
  physical validity in molecular simulations},}\ }\href@noop {} {\bibfield
  {journal} {\bibinfo  {journal} {PloS ONE}\ }\textbf {\bibinfo {volume}
  {13}},\ \bibinfo {pages} {e0202764} (\bibinfo {year} {2018})}\BibitemShut
  {NoStop}%
\bibitem [{\citenamefont {Shirts}\ and\ \citenamefont
  {Chodera}(2008)}]{shirts2008statistically}%
  \BibitemOpen
  \bibfield  {author} {\bibinfo {author} {\bibfnamefont {M.~R.}\ \bibnamefont
  {Shirts}}\ and\ \bibinfo {author} {\bibfnamefont {J.~D.}\ \bibnamefont
  {Chodera}},\ }\bibfield  {title} {\enquote {\bibinfo {title} {Statistically
  optimal analysis of samples from multiple equilibrium states},}\ }\href@noop
  {} {\bibfield  {journal} {\bibinfo  {journal} {J. Chem. Phys.}\ }\textbf
  {\bibinfo {volume} {129}},\ \bibinfo {pages} {124105} (\bibinfo {year}
  {2008})}\BibitemShut {NoStop}%
\bibitem [{\citenamefont {Klimovich}, \citenamefont {Shirts},\ and\
  \citenamefont {Mobley}(2015)}]{Klimovich:2015er}%
  \BibitemOpen
  \bibfield  {author} {\bibinfo {author} {\bibfnamefont {P.~V.}\ \bibnamefont
  {Klimovich}}, \bibinfo {author} {\bibfnamefont {M.~R.}\ \bibnamefont
  {Shirts}}, \ and\ \bibinfo {author} {\bibfnamefont {D.~L.}\ \bibnamefont
  {Mobley}},\ }\bibfield  {title} {\enquote {\bibinfo {title} {Guidelines for
  the analysis of free energy calculations},}\ }\href {\doibase
  10.1007/s10822-015-9840-9} {\bibfield  {journal} {\bibinfo  {journal} {J.
  Comput. Aided Mol. Des.}\ }\textbf {\bibinfo {volume} {29}},\ \bibinfo
  {pages} {397--411} (\bibinfo {year} {2015})}\BibitemShut {NoStop}%
\bibitem [{\citenamefont {Bennett}(1976)}]{bennett1976efficient}%
  \BibitemOpen
  \bibfield  {author} {\bibinfo {author} {\bibfnamefont {C.~H.}\ \bibnamefont
  {Bennett}},\ }\bibfield  {title} {\enquote {\bibinfo {title} {Efficient
  estimation of free energy differences from {M}onte {C}arlo data},}\
  }\href@noop {} {\bibfield  {journal} {\bibinfo  {journal} {J. Comput. Phys.}\
  }\textbf {\bibinfo {volume} {22}},\ \bibinfo {pages} {245--268} (\bibinfo
  {year} {1976})}\BibitemShut {NoStop}%
\bibitem [{\citenamefont {Wang}\ \emph {et~al.}(2004)\citenamefont {Wang},
  \citenamefont {Wolf}, \citenamefont {Caldwell}, \citenamefont {Kollman},\
  and\ \citenamefont {Case}}]{wang2004development}%
  \BibitemOpen
  \bibfield  {author} {\bibinfo {author} {\bibfnamefont {J.}~\bibnamefont
  {Wang}}, \bibinfo {author} {\bibfnamefont {R.~M.}\ \bibnamefont {Wolf}},
  \bibinfo {author} {\bibfnamefont {J.~W.}\ \bibnamefont {Caldwell}}, \bibinfo
  {author} {\bibfnamefont {P.~A.}\ \bibnamefont {Kollman}}, \ and\ \bibinfo
  {author} {\bibfnamefont {D.~A.}\ \bibnamefont {Case}},\ }\bibfield  {title}
  {\enquote {\bibinfo {title} {Development and testing of a general amber force
  field},}\ }\href@noop {} {\bibfield  {journal} {\bibinfo  {journal} {J.
  Comput. Chem.}\ }\textbf {\bibinfo {volume} {25}},\ \bibinfo {pages}
  {1157--1174} (\bibinfo {year} {2004})}\BibitemShut {NoStop}%
\bibitem [{\citenamefont {Bayly}\ \emph {et~al.}(1993)\citenamefont {Bayly},
  \citenamefont {Cieplak}, \citenamefont {Cornell},\ and\ \citenamefont
  {Kollman}}]{bayly1993well}%
  \BibitemOpen
  \bibfield  {author} {\bibinfo {author} {\bibfnamefont {C.~I.}\ \bibnamefont
  {Bayly}}, \bibinfo {author} {\bibfnamefont {P.}~\bibnamefont {Cieplak}},
  \bibinfo {author} {\bibfnamefont {W.}~\bibnamefont {Cornell}}, \ and\
  \bibinfo {author} {\bibfnamefont {P.~A.}\ \bibnamefont {Kollman}},\
  }\bibfield  {title} {\enquote {\bibinfo {title} {A well-behaved electrostatic
  potential based method using charge restraints for deriving atomic charges:
  the {RESP} model},}\ }\href@noop {} {\bibfield  {journal} {\bibinfo
  {journal} {J. Phys. Chem.}\ }\textbf {\bibinfo {volume} {97}},\ \bibinfo
  {pages} {10269--10280} (\bibinfo {year} {1993})}\BibitemShut {NoStop}%
\bibitem [{\citenamefont {Tribello}\ \emph {et~al.}(2014)\citenamefont
  {Tribello}, \citenamefont {Bonomi}, \citenamefont {Branduardi}, \citenamefont
  {Camilloni},\ and\ \citenamefont {Bussi}}]{tribello2014plumed}%
  \BibitemOpen
  \bibfield  {author} {\bibinfo {author} {\bibfnamefont {G.~A.}\ \bibnamefont
  {Tribello}}, \bibinfo {author} {\bibfnamefont {M.}~\bibnamefont {Bonomi}},
  \bibinfo {author} {\bibfnamefont {D.}~\bibnamefont {Branduardi}}, \bibinfo
  {author} {\bibfnamefont {C.}~\bibnamefont {Camilloni}}, \ and\ \bibinfo
  {author} {\bibfnamefont {G.}~\bibnamefont {Bussi}},\ }\bibfield  {title}
  {\enquote {\bibinfo {title} {{PLUMED} 2: New feathers for an old bird},}\
  }\href@noop {} {\bibfield  {journal} {\bibinfo  {journal} {Comput. Phys.
  Commun.}\ }\textbf {\bibinfo {volume} {185}},\ \bibinfo {pages} {604--613}
  (\bibinfo {year} {2014})}\BibitemShut {NoStop}%
\bibitem [{\citenamefont {Lemkul}(2018)}]{lemkul2018proteins}%
  \BibitemOpen
  \bibfield  {author} {\bibinfo {author} {\bibfnamefont {J.}~\bibnamefont
  {Lemkul}},\ }\bibfield  {title} {\enquote {\bibinfo {title} {From proteins to
  perturbed {H}amiltonians: A suite of tutorials for the {GROMACS}-2018
  molecular simulation package [article v1. 0]},}\ }\href@noop {} {\bibfield
  {journal} {\bibinfo  {journal} {Living J. Comp. Mol. Sci.}\ }\textbf
  {\bibinfo {volume} {1}},\ \bibinfo {pages} {5068} (\bibinfo {year}
  {2018})}\BibitemShut {NoStop}%
\bibitem [{\citenamefont {Kandasamy}\ and\ \citenamefont
  {Larson}(2006)}]{kandasamy2006molecular}%
  \BibitemOpen
  \bibfield  {author} {\bibinfo {author} {\bibfnamefont {S.~K.}\ \bibnamefont
  {Kandasamy}}\ and\ \bibinfo {author} {\bibfnamefont {R.~G.}\ \bibnamefont
  {Larson}},\ }\bibfield  {title} {\enquote {\bibinfo {title} {Molecular
  dynamics simulations of model trans-membrane peptides in lipid bilayers: a
  systematic investigation of hydrophobic mismatch},}\ }\href@noop {}
  {\bibfield  {journal} {\bibinfo  {journal} {Biophys. J.}\ }\textbf {\bibinfo
  {volume} {90}},\ \bibinfo {pages} {2326--2343} (\bibinfo {year}
  {2006})}\BibitemShut {NoStop}%
\bibitem [{\citenamefont {Oostenbrink}\ \emph {et~al.}(2004)\citenamefont
  {Oostenbrink}, \citenamefont {Villa}, \citenamefont {Mark},\ and\
  \citenamefont {Van~Gunsteren}}]{oostenbrink2004biomolecular}%
  \BibitemOpen
  \bibfield  {author} {\bibinfo {author} {\bibfnamefont {C.}~\bibnamefont
  {Oostenbrink}}, \bibinfo {author} {\bibfnamefont {A.}~\bibnamefont {Villa}},
  \bibinfo {author} {\bibfnamefont {A.~E.}\ \bibnamefont {Mark}}, \ and\
  \bibinfo {author} {\bibfnamefont {W.~F.}\ \bibnamefont {Van~Gunsteren}},\
  }\bibfield  {title} {\enquote {\bibinfo {title} {A biomolecular force field
  based on the free enthalpy of hydration and solvation: the {GROMOS}
  force-field parameter sets {53A5} and {53A6}},}\ }\href@noop {} {\bibfield
  {journal} {\bibinfo  {journal} {J. Comput. Chem.}\ }\textbf {\bibinfo
  {volume} {25}},\ \bibinfo {pages} {1656--1676} (\bibinfo {year}
  {2004})}\BibitemShut {NoStop}%
\bibitem [{\citenamefont {Berger}, \citenamefont {Edholm},\ and\ \citenamefont
  {J{\"a}hnig}(1997)}]{berger1997molecular}%
  \BibitemOpen
  \bibfield  {author} {\bibinfo {author} {\bibfnamefont {O.}~\bibnamefont
  {Berger}}, \bibinfo {author} {\bibfnamefont {O.}~\bibnamefont {Edholm}}, \
  and\ \bibinfo {author} {\bibfnamefont {F.}~\bibnamefont {J{\"a}hnig}},\
  }\bibfield  {title} {\enquote {\bibinfo {title} {Molecular dynamics
  simulations of a fluid bilayer of dipalmitoylphosphatidylcholine at full
  hydration, constant pressure, and constant temperature},}\ }\href@noop {}
  {\bibfield  {journal} {\bibinfo  {journal} {Biophys. J.}\ }\textbf {\bibinfo
  {volume} {72}},\ \bibinfo {pages} {2002--2013} (\bibinfo {year}
  {1997})}\BibitemShut {NoStop}%
\bibitem [{\citenamefont {Kandt}, \citenamefont {Ash},\ and\ \citenamefont
  {Tieleman}(2007)}]{kandt2007setting}%
  \BibitemOpen
  \bibfield  {author} {\bibinfo {author} {\bibfnamefont {C.}~\bibnamefont
  {Kandt}}, \bibinfo {author} {\bibfnamefont {W.~L.}\ \bibnamefont {Ash}}, \
  and\ \bibinfo {author} {\bibfnamefont {D.~P.}\ \bibnamefont {Tieleman}},\
  }\bibfield  {title} {\enquote {\bibinfo {title} {Setting up and running
  molecular dynamics simulations of membrane proteins},}\ }\href@noop {}
  {\bibfield  {journal} {\bibinfo  {journal} {Methods}\ }\textbf {\bibinfo
  {volume} {41}},\ \bibinfo {pages} {475--488} (\bibinfo {year}
  {2007})}\BibitemShut {NoStop}%
\bibitem [{\citenamefont {Berendsen}\ \emph {et~al.}(1981)\citenamefont
  {Berendsen}, \citenamefont {Postma}, \citenamefont {van Gunsteren},\ and\
  \citenamefont {Hermans}}]{berendsen1981interaction}%
  \BibitemOpen
  \bibfield  {author} {\bibinfo {author} {\bibfnamefont {H.~J.}\ \bibnamefont
  {Berendsen}}, \bibinfo {author} {\bibfnamefont {J.~P.}\ \bibnamefont
  {Postma}}, \bibinfo {author} {\bibfnamefont {W.~F.}\ \bibnamefont {van
  Gunsteren}}, \ and\ \bibinfo {author} {\bibfnamefont {J.}~\bibnamefont
  {Hermans}},\ }\bibfield  {title} {\enquote {\bibinfo {title} {Interaction
  models for water in relation to protein hydration},}\ }in\ \href@noop {}
  {\emph {\bibinfo {booktitle} {Intermolecular forces}}}\ (\bibinfo
  {publisher} {Springer},\ \bibinfo {year} {1981})\ pp.\ \bibinfo {pages}
  {331--342}\BibitemShut {NoStop}%
\bibitem [{\citenamefont {Nos{\'e}}(1984)}]{nose1984unified}%
  \BibitemOpen
  \bibfield  {author} {\bibinfo {author} {\bibfnamefont {S.}~\bibnamefont
  {Nos{\'e}}},\ }\bibfield  {title} {\enquote {\bibinfo {title} {A unified
  formulation of the constant temperature molecular dynamics methods},}\
  }\href@noop {} {\bibfield  {journal} {\bibinfo  {journal} {J. Chem. Phys.}\
  }\textbf {\bibinfo {volume} {81}},\ \bibinfo {pages} {511--519} (\bibinfo
  {year} {1984})}\BibitemShut {NoStop}%
\bibitem [{\citenamefont {Flyvbjerg}\ and\ \citenamefont
  {Petersen}(1989)}]{flyvbjerg1989error}%
  \BibitemOpen
  \bibfield  {author} {\bibinfo {author} {\bibfnamefont {H.}~\bibnamefont
  {Flyvbjerg}}\ and\ \bibinfo {author} {\bibfnamefont {H.~G.}\ \bibnamefont
  {Petersen}},\ }\bibfield  {title} {\enquote {\bibinfo {title} {Error
  estimates on averages of correlated data},}\ }\href@noop {} {\bibfield
  {journal} {\bibinfo  {journal} {J. Chem. Phys.}\ }\textbf {\bibinfo {volume}
  {91}},\ \bibinfo {pages} {461--466} (\bibinfo {year} {1989})}\BibitemShut
  {NoStop}%
\bibitem [{\citenamefont {Jorgensen}\ and\ \citenamefont
  {Jenson}(1998)}]{jorgensen1998temperature}%
  \BibitemOpen
  \bibfield  {author} {\bibinfo {author} {\bibfnamefont {W.~L.}\ \bibnamefont
  {Jorgensen}}\ and\ \bibinfo {author} {\bibfnamefont {C.}~\bibnamefont
  {Jenson}},\ }\bibfield  {title} {\enquote {\bibinfo {title} {Temperature
  dependence of {TIP3P}, {SPC}, and {TIP4P} water from {NPT} {M}onte {C}arlo
  simulations: Seeking temperatures of maximum density},}\ }\href@noop {}
  {\bibfield  {journal} {\bibinfo  {journal} {J. Comput. Chem.}\ }\textbf
  {\bibinfo {volume} {19}},\ \bibinfo {pages} {1179--1186} (\bibinfo {year}
  {1998})}\BibitemShut {NoStop}%
\bibitem [{\citenamefont {Schneider}, \citenamefont {Stoll},\ and\
  \citenamefont {Morf}(1978)}]{schneider1978brownian}%
  \BibitemOpen
  \bibfield  {author} {\bibinfo {author} {\bibfnamefont {T.}~\bibnamefont
  {Schneider}}, \bibinfo {author} {\bibfnamefont {E.}~\bibnamefont {Stoll}}, \
  and\ \bibinfo {author} {\bibfnamefont {R.}~\bibnamefont {Morf}},\ }\bibfield
  {title} {\enquote {\bibinfo {title} {Brownian motion of interacting and
  noninteracting particles subject to a periodic potential and driven by an
  external field},}\ }\href@noop {} {\bibfield  {journal} {\bibinfo  {journal}
  {Phys. Rev. B}\ }\textbf {\bibinfo {volume} {18}},\ \bibinfo {pages} {1417}
  (\bibinfo {year} {1978})}\BibitemShut {NoStop}%
\bibitem [{\citenamefont {Bussi}\ and\ \citenamefont
  {Parrinello}(2008)}]{bussi2008stochastic}%
  \BibitemOpen
  \bibfield  {author} {\bibinfo {author} {\bibfnamefont {G.}~\bibnamefont
  {Bussi}}\ and\ \bibinfo {author} {\bibfnamefont {M.}~\bibnamefont
  {Parrinello}},\ }\bibfield  {title} {\enquote {\bibinfo {title} {Stochastic
  thermostats: comparison of local and global schemes},}\ }\href@noop {}
  {\bibfield  {journal} {\bibinfo  {journal} {Comput. Phys. Commun.}\ }\textbf
  {\bibinfo {volume} {179}},\ \bibinfo {pages} {26--29} (\bibinfo {year}
  {2008})}\BibitemShut {NoStop}%
\bibitem [{\citenamefont {Harvey}, \citenamefont {Tan},\ and\ \citenamefont
  {Cheatham~III}(1998)}]{harvey1998flying}%
  \BibitemOpen
  \bibfield  {author} {\bibinfo {author} {\bibfnamefont {S.~C.}\ \bibnamefont
  {Harvey}}, \bibinfo {author} {\bibfnamefont {R.~K.-Z.}\ \bibnamefont {Tan}},
  \ and\ \bibinfo {author} {\bibfnamefont {T.~E.}\ \bibnamefont
  {Cheatham~III}},\ }\bibfield  {title} {\enquote {\bibinfo {title} {The flying
  ice cube: velocity rescaling in molecular dynamics leads to violation of
  energy equipartition},}\ }\href@noop {} {\bibfield  {journal} {\bibinfo
  {journal} {J. Comput. Chem.}\ }\textbf {\bibinfo {volume} {19}},\ \bibinfo
  {pages} {726--740} (\bibinfo {year} {1998})}\BibitemShut {NoStop}%
\bibitem [{\citenamefont {Rosta}, \citenamefont {Buchete},\ and\ \citenamefont
  {Hummer}(2009)}]{rosta2009thermostat}%
  \BibitemOpen
  \bibfield  {author} {\bibinfo {author} {\bibfnamefont {E.}~\bibnamefont
  {Rosta}}, \bibinfo {author} {\bibfnamefont {N.-V.}\ \bibnamefont {Buchete}},
  \ and\ \bibinfo {author} {\bibfnamefont {G.}~\bibnamefont {Hummer}},\
  }\bibfield  {title} {\enquote {\bibinfo {title} {Thermostat artifacts in
  replica exchange molecular dynamics simulations},}\ }\href@noop {} {\bibfield
   {journal} {\bibinfo  {journal} {J. Chem. Theory Comput.}\ }\textbf {\bibinfo
  {volume} {5}},\ \bibinfo {pages} {1393--1399} (\bibinfo {year}
  {2009})}\BibitemShut {NoStop}%
\bibitem [{\citenamefont {Wong-Ekkabut}\ \emph {et~al.}(2010)\citenamefont
  {Wong-Ekkabut}, \citenamefont {Miettinen}, \citenamefont {Dias},\ and\
  \citenamefont {Karttunen}}]{wong2010static}%
  \BibitemOpen
  \bibfield  {author} {\bibinfo {author} {\bibfnamefont {J.}~\bibnamefont
  {Wong-Ekkabut}}, \bibinfo {author} {\bibfnamefont {M.~S.}\ \bibnamefont
  {Miettinen}}, \bibinfo {author} {\bibfnamefont {C.}~\bibnamefont {Dias}}, \
  and\ \bibinfo {author} {\bibfnamefont {M.}~\bibnamefont {Karttunen}},\
  }\bibfield  {title} {\enquote {\bibinfo {title} {Static charges cannot drive
  a continuous flow of water molecules through a carbon nanotube},}\
  }\href@noop {} {\bibfield  {journal} {\bibinfo  {journal} {Nat.
  Nanotechnol.}\ }\textbf {\bibinfo {volume} {5}},\ \bibinfo {pages} {555--557}
  (\bibinfo {year} {2010})}\BibitemShut {NoStop}%
\bibitem [{\citenamefont {Wong-ekkabut}\ and\ \citenamefont
  {Karttunen}(2016)}]{wong2016good}%
  \BibitemOpen
  \bibfield  {author} {\bibinfo {author} {\bibfnamefont {J.}~\bibnamefont
  {Wong-ekkabut}}\ and\ \bibinfo {author} {\bibfnamefont {M.}~\bibnamefont
  {Karttunen}},\ }\bibfield  {title} {\enquote {\bibinfo {title} {The good, the
  bad and the user in soft matter simulations},}\ }\href@noop {} {\bibfield
  {journal} {\bibinfo  {journal} {Biochim. Biophys. Acta-Biomembranes}\
  }\textbf {\bibinfo {volume} {1858}},\ \bibinfo {pages} {2529--2538} (\bibinfo
  {year} {2016})}\BibitemShut {NoStop}%
\bibitem [{\citenamefont {Rogge}\ \emph {et~al.}(2015)\citenamefont {Rogge},
  \citenamefont {Vanduyfhuys}, \citenamefont {Ghysels}, \citenamefont
  {Waroquier}, \citenamefont {Verstraelen}, \citenamefont {Maurin},\ and\
  \citenamefont {Van~Speybroeck}}]{rogge2015comparison}%
  \BibitemOpen
  \bibfield  {author} {\bibinfo {author} {\bibfnamefont {S.}~\bibnamefont
  {Rogge}}, \bibinfo {author} {\bibfnamefont {L.}~\bibnamefont {Vanduyfhuys}},
  \bibinfo {author} {\bibfnamefont {A.}~\bibnamefont {Ghysels}}, \bibinfo
  {author} {\bibfnamefont {M.}~\bibnamefont {Waroquier}}, \bibinfo {author}
  {\bibfnamefont {T.}~\bibnamefont {Verstraelen}}, \bibinfo {author}
  {\bibfnamefont {G.}~\bibnamefont {Maurin}}, \ and\ \bibinfo {author}
  {\bibfnamefont {V.}~\bibnamefont {Van~Speybroeck}},\ }\bibfield  {title}
  {\enquote {\bibinfo {title} {A comparison of barostats for the mechanical
  characterization of metal--organic frameworks},}\ }\href@noop {} {\bibfield
  {journal} {\bibinfo  {journal} {J. Chem. Theory Comput.}\ }\textbf {\bibinfo
  {volume} {11}},\ \bibinfo {pages} {5583--5597} (\bibinfo {year}
  {2015})}\BibitemShut {NoStop}%
\bibitem [{\citenamefont {Okabe}\ \emph {et~al.}(2001)\citenamefont {Okabe},
  \citenamefont {Kawata}, \citenamefont {Okamoto},\ and\ \citenamefont
  {Mikami}}]{okabe2001replica}%
  \BibitemOpen
  \bibfield  {author} {\bibinfo {author} {\bibfnamefont {T.}~\bibnamefont
  {Okabe}}, \bibinfo {author} {\bibfnamefont {M.}~\bibnamefont {Kawata}},
  \bibinfo {author} {\bibfnamefont {Y.}~\bibnamefont {Okamoto}}, \ and\
  \bibinfo {author} {\bibfnamefont {M.}~\bibnamefont {Mikami}},\ }\bibfield
  {title} {\enquote {\bibinfo {title} {Replica-exchange monte carlo method for
  the isobaric--isothermal ensemble},}\ }\href@noop {} {\bibfield  {journal}
  {\bibinfo  {journal} {Chem. Phys. Lett.}\ }\textbf {\bibinfo {volume}
  {335}},\ \bibinfo {pages} {435--439} (\bibinfo {year} {2001})}\BibitemShut
  {NoStop}%
\bibitem [{\citenamefont {Mori}, \citenamefont {Jung},\ and\ \citenamefont
  {Sugita}(2013)}]{mori2013surface}%
  \BibitemOpen
  \bibfield  {author} {\bibinfo {author} {\bibfnamefont {T.}~\bibnamefont
  {Mori}}, \bibinfo {author} {\bibfnamefont {J.}~\bibnamefont {Jung}}, \ and\
  \bibinfo {author} {\bibfnamefont {Y.}~\bibnamefont {Sugita}},\ }\bibfield
  {title} {\enquote {\bibinfo {title} {Surface-tension replica-exchange
  molecular dynamics method for enhanced sampling of biological membrane
  systems},}\ }\href@noop {} {\bibfield  {journal} {\bibinfo  {journal} {J.
  Chem. Theory Comput.}\ }\textbf {\bibinfo {volume} {9}},\ \bibinfo {pages}
  {5629--5640} (\bibinfo {year} {2013})}\BibitemShut {NoStop}%
\end{thebibliography}%

\end{document}